# Basics of RF Electronics

*A.C. Dexter*
Cockcroft Institute – Lancaster University, Lancaster, UK

**Abstract**
The focus of this educational text is selected examples of high-frequency electronic circuits and their components employed for the accurate phasing and synchronisation of accelerator cavities. Examples have been chosen to describe the basics of RF electronics. The starting point is transmission lines, connectors, discontinuities, and the handling of reflection. The application of simple surface mount components is discussed. The use of the Kuroda identities for converting lumped circuit designs to printed circuit designs is demonstrated. The accelerator example used to demonstrate the use of components is a circuit designed for the synchronisation of the CLIC crab cavities. This example employs co-planar waveguide, SMA connectors, Wilkinson splitters, and surface-mount double-balanced mixers. For the control of cavity phase and amplitude, the benefit of I&Q controllers will be explained. The text will then discuss the operation and use of I&Q modulators and VCOs.

## 1    Introduction

As this text is part of a course, it assumes knowledge of material that has previously been presented but holds back on topics that are yet to come and more specifically, Low-Level Radio Frequency (LLRF). The material presented here is a very small subset of RF electronics which includes circuits and components for communications systems, radar, imaging, RF measurement and control systems for particle accelerators. The topic "Basics of RF electronics" has been delivered previously at the 2010 CAS "RF for accelerators". CAS texts have been archived and are available to all [1], hence we avoid following the previous approach by S. Gallo [2] too closely but recommend that you study his text in detail if this subject is important for your research. Gallo discusses a very wide range of components. Here a much smaller range of components is discussed and there is more focus on fundamentals.

When electronics is taught, a distinction is often made between alternating current theory (AC), radio frequency engineering (RF) and distributed circuits. For AC theory the assumption is made that transmission line theory is not needed to model electrical connections between components and hence the reflection of power by discontinuities does not arise. In this text, we start by revising transmission line theory and relate this to the selection, application and mounting of PCB components. Detailed consideration is given to reflection and how it should be analysed. Some printed PCB structures are briefly discussed. The section on filters is minimal on their fundamental design but covers the Kuroda identities which are needed if one wishes to implement one's own distributed filter on a PCB.

The measurement of cavity phase is critical for accelerator complexes with multiple RF power amplifiers. Gallo describes several methods of measuring phase and several types of mixer. For reason of space here the focus is on the double-balanced mixer. One approach to LLRF systems to control phase and amplitude in a cavity is I&Q control. In this text, we will explain the advantage of I&Q control. The basic theory of oscillators will be covered with circuit examples and applications. Finally, there are sections on the basic theory of phase noise and phase-locked loops.

## 2  Transmission line theory

### 2.1  Waves on lossless lines

A distinction between AC theory and RF electronics is the need to consider many circuit interconnects as transmission lines. The simplest transmission line is two parallel, and relatively straight wires connecting a generator to a load, as shown in Fig.1. Prior to the switch being closed, there will be an electric field across the generator terminals and therefore an electric field, and hence a voltage V across the open switch. Assuming the switch has only a tiny capacitance $C_s$, then only a tiny charge Q is needed to establish the voltage ($Q = C_sV$). At the instant the switch is closed, the electric field, which was across the switch, now drives current along the wire. At this instant the generator has no knowledge of the load but sees the capacitance between the conductors close to the switch, hence charge must flow to raise this additional capacitance to the voltage of the generator. Solving Maxwell's equations, one sees the wavefront of a transverse electric and magnetic field mode (TEM) propagating between and in the direction of the wires at the velocity of light for the material that separates the conductors.

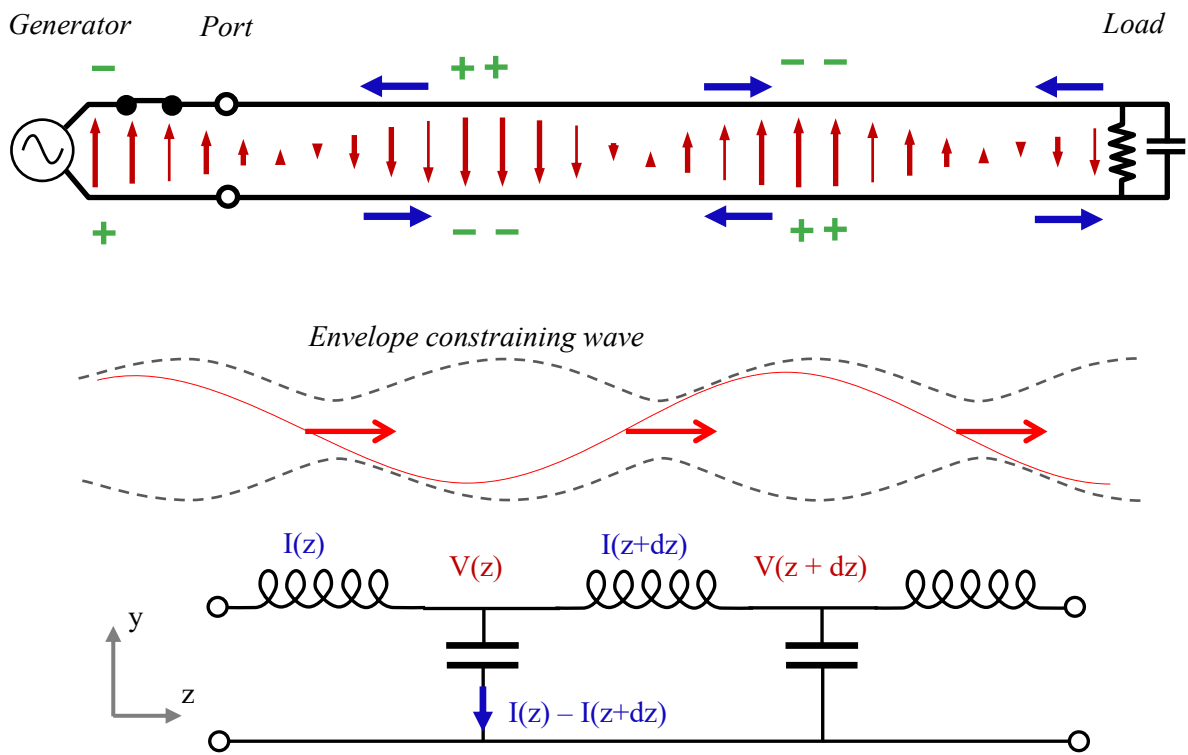

**Fig. 1:** *Top:* Transmission line composed of parallel conductors showing fields in red, currents in blue and charge in green. *Middle:* In grey is the standing wave envelope through which the electric field wave waxes and wanes when there is some reflection from the load. *Bottom*: is the equivalent circuit.

Naively one might think that if the wavefront moves at the velocity of light, then the electrons providing an excess or deficit of charge must also move at the speed of light. This is not the case as the electrons in the metal move very slowly. A very large number of electrons moving slowly from their nuclear positive cores gives a very large transverse electric field and it is this electric field that propagates at the velocity of light. As it propagates it shifts new electrons further along the line. Behind the wavefront is a travelling wave. The ratio between the peak voltage between the conductors and the peak current flowing along the conductors is the impedance. If energy only flows in one direction (a forward wave) then the current will be in phase with the voltage and hence the impedance is real. If the impedance of the load differs from the line, then there will be reflection. Total reflection gives a standing wave and the current is out of phase with the voltage. For partial reflection, the wave waxes and wanes as it moves through an envelope as shown in the middle diagram of Fig. 1. The ratio of the envelope maximum to the envelope minimum is known as the voltage standing wave ratio (VSWR). Before the availability of Network analysers RF engineers measured the VSWR with a probe.

Rather than solving Maxwell's equations, it is more convenient to solve the problem with circuit equations. The equivalent circuit assuming no losses is given in the lower diagram of Fig. 1. Applying the circuit equation "*voltage difference equals inductance times rate of change of current along the conductor*" and "*current between the conductors is capacitance times rate of change of voltage*" gives the Eqs. (1) and (2),

$$\frac{\partial V}{\partial z} = -L \frac{\partial I}{\partial t} \qquad (1) \qquad\qquad \frac{\partial I}{\partial z} = -C \frac{\partial V}{\partial t}, \qquad (2)$$

where L is the inductance per unit length and C is the capacitance per unit length. Solving gives the wave equation $\frac{\partial^2 V}{\partial z^2} = LC \frac{\partial^2 V}{\partial t^2}$ with phase velocity $c = \frac{1}{\sqrt{LC}}$. Since the velocity of the wave only depends on the material between the lines, one realises that once the capacitance per unit length has been determined then so has the inductance per unit length. The situation is more complex if one has a mix of materials as one has for microstrip and co-planar waveguide etc. see next section.

Solutions of the wave equation can be decomposed into forward and backwards waves as

$$V(z,t) = F\left(\frac{t}{\sqrt{LC}} - z\right) + R\left(\frac{t}{\sqrt{LC}} + z\right), \qquad (3)$$

where F is the information signal travelling to the right and R is the information signal travelling to the left. The functional forms of F and R determine the information and power density being sent or returned. Application of either Eqs. (1) or (2) to (3) gives,

$$I(z,t) = \sqrt{\frac{C}{L}} \left\{ F\left(\frac{t}{\sqrt{LC}} - z\right) - R\left(\frac{t}{\sqrt{LC}} + z\right) \right\}. \qquad (4)$$

Considering the ratio of voltage to current for a forward wave with no reflection gives the intrinsic impedance as $Z_o = \sqrt{\frac{L}{C}}$. If the voltage and current are known at a point, then Eqs. (1) and (2) allow the forward and backward waves to be determined. As the measurement of voltage and current at high frequencies is difficult, RF engineers measure and calculate with the forward and reflected waves hence the importance of scattering matrices.

Figure 1 marks a port. When considering RF circuits, it is convenient to consider sections independently. As the phase on the earth depends on location, one cannot complete circuit diagrams with arbitrary connections to earth. Sections of RF circuits are defined by pairs of terminals forming a **port**. The condition for a pair of terminals to constitute a port is that the current entering one, equals the current leaving the other, in both phase and amplitude.

For connections between LLRF circuits on accelerators, one invariably chooses co-axial cable with an intrinsic impedance of 50 Ω. This value was adopted as a standard for instrumentation, being a compromise between minimising attenuation and maximising power transfer, Z. Peterson [3]. The power handling can be increased by increasing the diameters of the conductors subject to one's ability to cool the inner conductor. The useful limit for measurement systems is the dimension where over-moding occurs at the frequency of interest. A standard SMA co-axial cable is rated for mode-free operation to 18 GHz [4], see Fig. 2. The cut-through image in Fig. 2 has been made to a printed circuit board (PCB) connector.

Individual components for an RF circuit are typically mounted onto a PCB. Tracks on the PCB would normally be designed to be 50 Ω to avoid reflection at the connector. The third image in Fig. 2 is a PCB edge connector to SMA.

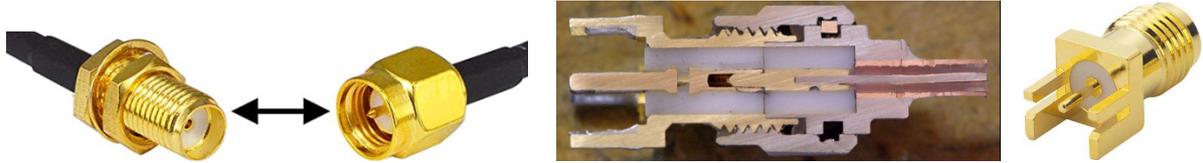

**Fig. 2:** SMA connectors and co-axial cable

## 2.2 Transmission line reflection

If it is necessary to measure phase at a location, for example, an accelerator cavity, then one might account for the transmission path back to the generator and derive the result from the phase at the generator. For high-accuracy measurements, the result can be affected by reflections along the path. Figure 3 illustrates this situation with two connectors on a transmission line. Specialising the general signal waveforms of Eqs. (3) and (4) to sinusoidal travelling waves with wave number k and using a complex number format, then waves going to the right and left respectively can be written in the form $F \exp j(kz - \omega t)$ and $R \exp j(-kz - \omega t)$ where F and R are now constant complex amplitudes. As all the waves have the same time dependence then the factor $\exp(-j\omega t)$ can be omitted from the analysis.

In Figure 3 there is a forward wave $F_1$ incident on the first connector, $R_1$ is reflected and $F_2$ is transmitted. The wave travelling from connector one to connector two will have gained a phase of ka when it arrives. The reflection is $R_2$ and as it travels distance -a before arriving back at connector 1, it also gains a phase of ka. Assuming the load is perfectly matched, the final leg only carries a forward wave $F_3$. The problem is to determine how the phase of wave $F_3$ varies with imperfect connectors.

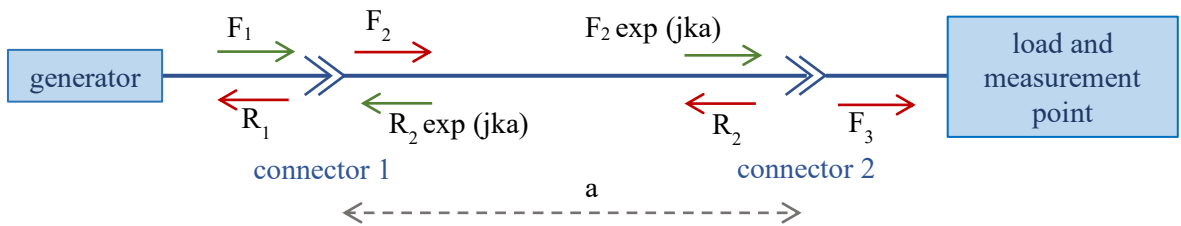

**Fig. 3:** Reflections at connectors on a transmission line.

Taking the S matrices for the connectors to be unitary, and dependent on a single variable providing a slight perturbation from the identity matrix, then the matrix equations for the system take the form:

$$\begin{bmatrix} R_1 \\ F_2 \end{bmatrix} = \begin{bmatrix} j\delta & \sqrt{1-\delta^2} \\ \sqrt{1-\delta^2} & j\delta \end{bmatrix} \begin{bmatrix} F_1 \\ R_2 \exp(jka) \end{bmatrix} \qquad \begin{bmatrix} R_2 \\ F_3 \end{bmatrix} = \begin{bmatrix} j\delta & \sqrt{1-\delta^2} \\ \sqrt{1-\delta^2} & j\delta \end{bmatrix} \begin{bmatrix} F_2 \exp(jka) \\ 0 \end{bmatrix}$$

where δ is real. Solving gives

$$F_3 = \frac{1-\delta^2}{1+\delta^2 \exp(2jka)} F_1 \exp(jka).$$

Ideal transmission when $\delta = 0$ is $F_3 = F_1 \exp(jka)$. When $\delta \neq 0$ the denominator of the leading factor is complex so gives a phase shift with respect to ideal transmission. For the special case of $2ka = (2n-1)\pi$ where n is an integer, transmission is ideal. When $ka = n\pi$ the phase will be as for the $\delta = 0$ case but the amplitude is reduced by a factor $\frac{1-\delta^2}{1+\delta^2}$.

By choosing all transmission line segments between connectors or other discontinuities to be an exact number of half wavelengths, then phase errors are eliminated. This precaution could be significant when femtosecond level synchronisation is required.

When there is no reflection on a transmission line, the ratio of the voltage to the current gives the intrinsic impedance $Z_o$ which is independent of frequency for a dispersion-less line. When a line has reflection, then the impedance, again given as a ratio of the voltage to the current, is a useful quantity for circuit analysis at a specified frequency and it will depend on location. Using expressions for the complex waveforms given above and also Eqs. (3) and (4) one obtains:

$$Z(z) = Z_o \frac{F\exp(jkz) + R\exp(-jkz)}{F\exp(jkz) - R\exp(-jkz)}. \tag{5}$$

## 2.3 Co-planar waveguide

Most of the LLRF components that are used in measurement and control are available as surface mount components. A PCB by its fabrication method has printed copper tracks on a substrate. Substrates with printed tracks are glued together to make multilayer boards where tracks are internal to the board. At high frequencies, the tracks must be designed as transmission lines. A co-axial cable resists against pickup of airborne electromagnetic interference as the inner conductor is surrounded by an earthed sheath. For a track on the surface of a board, it is usually inconvenient to cover it with an additional earthed layer. A type of transmission line that is widely used with PCBs is grounded co-planar waveguide (GCPW) as shown in Fig. 4.

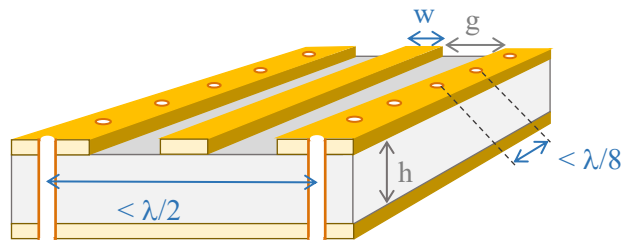

**Fig. 4:** Co-planar waveguide

The difference between microstrip and GCPW is ground planes on either side of the strip. By varying the gap between the strip and the adjacent earth planes, the impedance can be adjusted over a small range that is independent of the strip width. This means that the strip width can be adjusted to the same width as surface mount components. Closely spaced vias (conducting paths) are used to connect the upper earth plane with the lower earth plane.

A good approximation for the impedance is obtained using the following equations [5]:

$$k_1 = w + 2g \qquad k_2 = \sqrt{1-k_1^2} \qquad k_3 = \frac{\tanh\{0.25\pi w/h\}}{\tanh\{0.25\pi(w+2g)/h\}} \qquad k_4 = \sqrt{1-k_3^2}$$

$$\varepsilon_{eff} = \frac{K(k_1)K(k_4) + \varepsilon_r K(k_2)K(k_3)}{K(k_1)K(k_4) + K(k_2)K(k_3)} \quad (6) \qquad Z_o = \frac{60\pi}{\sqrt{\varepsilon_{eff}}} \frac{K(k_2)K(k_4)}{K(k_1)K(k_4) + K(k_2)K(k_3)}, \quad (7)$$

where K(k) is the complete elliptic integral of the first kind, w is the track width, g is the gap, and h is the substrate thickness. For frequencies up to a few GHz the substrate referred to as FR-4 is most commonly used. This is a flame retardant board made from woven glass fibre, epoxy resin and a mineral filler. Its high-frequency properties are not tightly controlled or specified, and above 6 GHz losses are usually unacceptable. Special low-loss substrates such as the Rogers RO4000 series [6] should be used at high frequencies. As an example, for RO4350B the manufacturers quote a loss factor of 0.0037 and a design relative permittivity (Dk) of 3.66 at 10 GHz. Fig. 5 evaluates the above formula for the GCPW track impedance of RO4350B with a laminate thickness of 0.506 mm as a function of track width and for differing gaps. Reading from the graph, one can form a 50 Ω track with a width of 0.85 mm when the gap is 0.2 mm.

The formula does not account for track thickness and indeed more accurate results would be given by direct simulation or using advanced design tools such as Altium PCB designer.

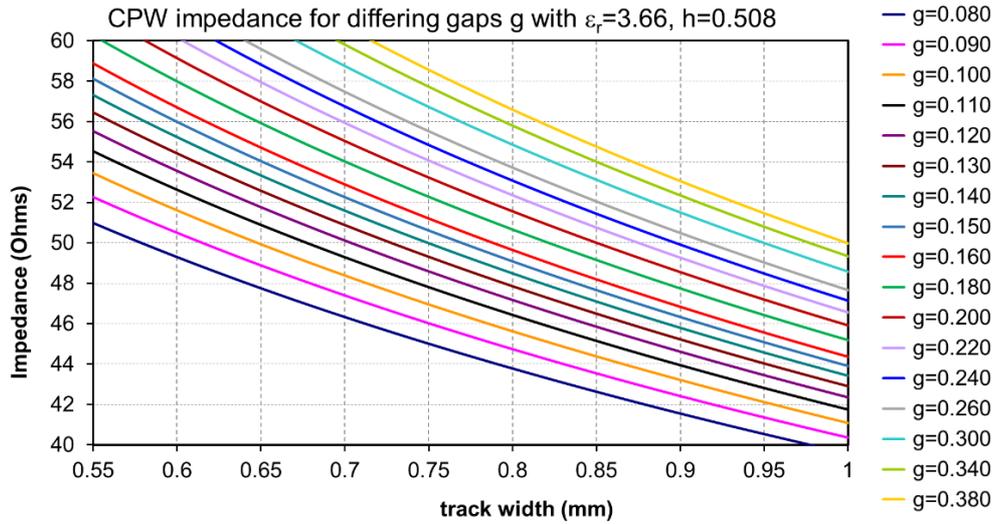

**Fig. 5:** Co-planar waveguide impedance as a function of track width for differing gaps from the track to the earth. Substrate thickness = 0.508 mm, permittivity = 3.66.

When performance is critical, test boards should be fabricated. Two were fabricated prior to designing a phase measurement PCB for the CLIC crab cavity synchronisation tests [7]. The crab cavities planned for the 3 TeV CLIC would be 40 metres apart, pulsed for just 200 ns but must be synchronised to 5 fs. The minimisation of reflection was fundamental to the design of the phase measurement board which forms part of an RF interferometer. The second test board is shown in Fig. 6 (*left*). The final phase measurement board is shown in Fig. 6 (*right*).

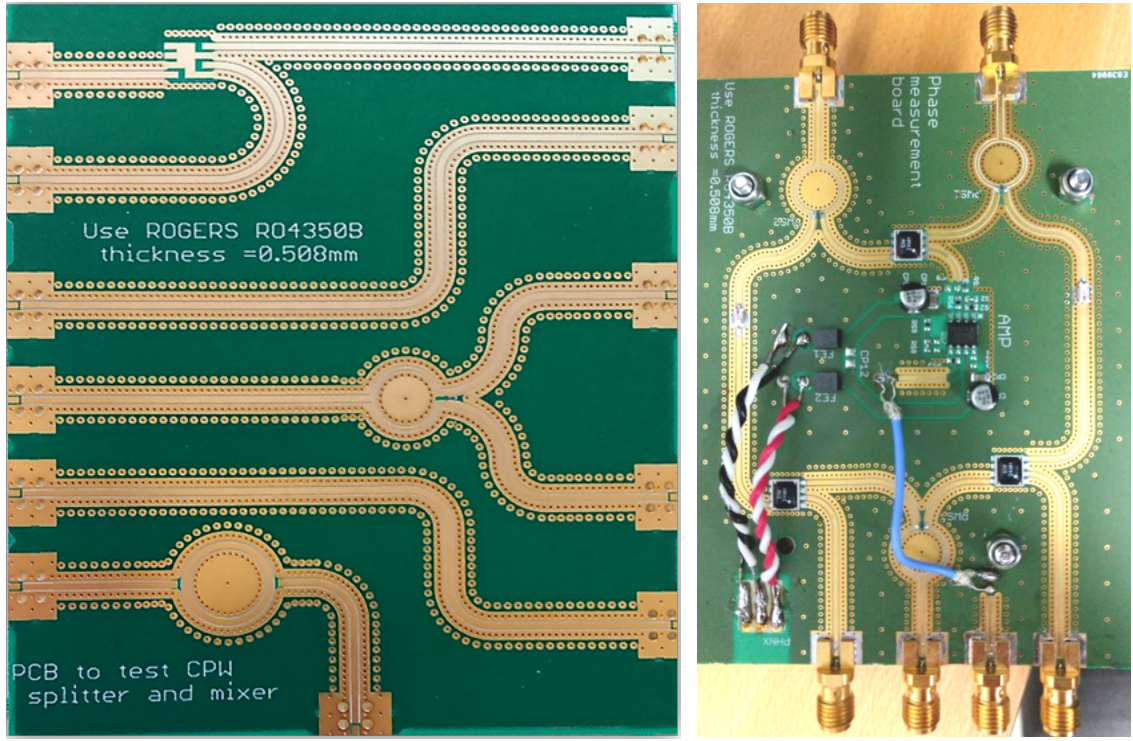

**Fig. 6:** (left)12 GHz test board for GCPW, (right) phase measurement board

The top and bottom surfaces are continuous 0.018 mm copper layers except for the thin gaps that define the tracks. The green covering is the solder mask. Exposed tracks and pads are gold-plated. The first test board allowed the solder mask to cover the gaps, and this was found to significantly change the track impedance.

The test board has five structures. The top structure has a footprint for the Mini-Circuits SIM-24MH+ mixer. The 2$^{nd}$ and 4$^{th}$ structures from the top are tracks which differ in length by 3.897 mm which should be exactly ¼ of a wavelength at the CLIC frequency of 12 GHz. The middle structure is for a Wilkinson splitter with ¾ wavelength loops. The bottom structure is a ring resonator with two capacitively coupled ports to the GCPW. Surface mount capacitors would cover the green solder mask patches to get good coupling.

After analysis of transmission for the tracks differing by ¼ wavelength on the test board, it was determined that the required design permittivity for this fabrication and using the formula above was close to 3.545 rather than 3.66. With this revised permittivity the required track width for a gap of 0.2 mm became 0.789 mm. Note that the wavelength for the GCPW is given as $\lambda = \lambda_o / \sqrt{\varepsilon_{eff}}$, where $\lambda_o$ is the free space wavelength and Eq. (6) gives the relationship between the effective permittivity $\varepsilon_{eff}$ for the GCPW air to dielectric boundary for a substrate with manufacturer's specified design permittivity $\varepsilon_r$. The wavelength for the GCPW as described at 12 GHz is 15.588 mm.

Both PCBs in Fig. 6 were two-layer boards with continuous ground planes. Most commercial PCBs have at least four layers. With multiple layers, one or more can be used to supply DC power. For the two-layer board shown, power has to be taken to the amplifier with ugly flying leads. A coaxial lead (blue) is used to take a low frequency (< 100 MHz) to an edge connector. A track width of 0.789 mm was convenient for the components used in the phase measurement board. High-speed data processing chips now commonly have a pin separation of 0.5 mm and pin width of 0.3 mm. In order to match the pin size to the track width one has to reduce the substrate thickness. Smaller thicknesses of 0.1 mm, 0.17 mm, and 0.25 mm are available for RO4350B but need to be used in a multilayer board to provide sufficient mechanical strength. The track width of 0.789 mm was perfect for the SIM-24MH+ mixer.

## 3 Surface mount components

Useful RF circuits typically need both active and passive components. Active components can amplify, rectify and generate new frequencies. The basic building blocks of active components are transistors and diodes, however, when building systems, one typically uses components with more sophisticated functionality like mixers and amplifiers, and these contain multiple building blocks. To keep these active components as small as possible and to maximise their range of operation, often their implementation requires external passive components, which include resistors, capacitors, and inductors.

The Wilkinson splitters shown in Fig. 6 require a 100 Ω resistor between the tracks at the fork. The operation of the splitter will be described in Section 6. Surface mount resistors, capacitors and inductors have standardised case sizes. The common reference is their length in thousandths of an inch, however, another common reference is their length in microns. The resistors used on the Wilkinson splitter and for biasing the operational amplifier are size "0402" (thousandths of an inch). Table 1 gives case size lengths and widths in millimetres. The electrical length for a series resistance will be significant with respect to the wavelength at GHz frequencies and hence might be a design consideration. One tries to match series components to the track width hence for a track width of 0.8 mm choose "0603". The Wilkinson splitter resistor is a shunt resistor that should be made as small as possible. One limitation is the necessary power handling, and another is the ease of construction. Students are typically able to mount 0603 components. Skilled technicians can mount 0402 components by hand. Robotic manipulators are usually needed to mount 0201 components. As one might expect, for the Wilkinson splitter the resistor needs to be mounted perfectly symmetrically to get a symmetric S-matrix.

**Table 1:** Passive component sizes taken from Vishey [8]

| Case length (1/1000 inch) | Case length (mm) | Case width (mm) | Gap between end caps (mm) |
|---|---|---|---|
| 0201 | 0.6 | 0.3 | 0.16 |
| 0402 | 1.005 | 0.5 | 0.5 |
| 0603 | 1.608 | 0.8 | 1.0 |
| 0805 | 2.012 | 1.25 | 1.3 |
| 1206 | 3.20 | 1.60 | 2.4 |

At high frequencies, a surface mount resistor may not function as a pure resistance. Its conducting path will have inductance even if it is a film and there will be capacitance between its terminals. A possible equivalent circuit is given in Fig. 7. R is the intended resistance. The red inductor and capacitor are parasitics. The capacitance to the earth plane is shown as the chip resistor is assumed to be on a PCB. If the resistor width is similar to the track width and the conducting path is close to the PCB substrate, then the resistor will to some degree, act as a lossy transmission line of a similar impedance to the track.

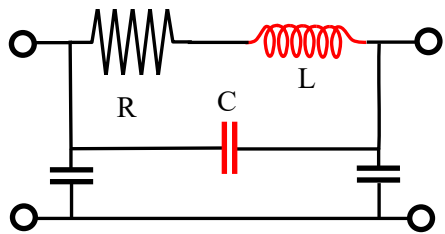

**Fig. 7:** High-frequency equivalent circuit for a film resistance

Surface mount capacitors are more problematic with respect to their high-frequency circuit behaviour. Firstly, one should choose a capacitor fabricated for high-frequency operation, having done this, one still has to recognise limits to its high-frequency response. For all but the very smallest capacitance values, surface mount capacitors will be made from multilayers and the gaps between the layers will be much smaller than the PCB substrate thickness hence the capacitance to earth on the ground plane can be neglected in an equivalent circuit. Fig. 8 (left) shows the physical structure and Fig. 8 (right) shows one possible simplified equivalent circuit. Importantly it is an equivalent circuit that supports a series and parallel resonance.

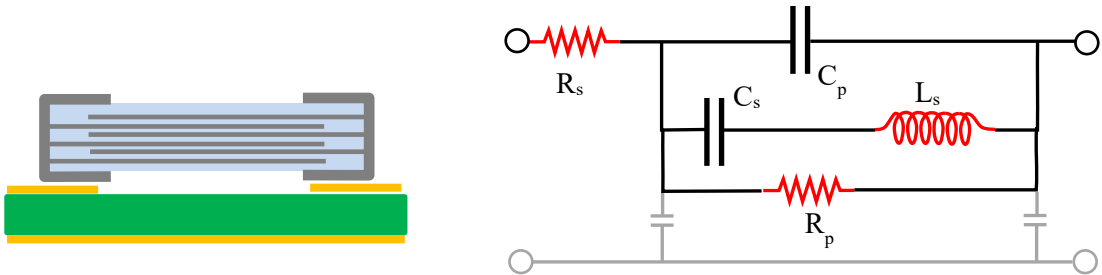

**Fig. 8:** Surface mount capacitor, physical structure (left) simplified high-frequency equivalent circuit (right).

The overall low-frequency capacitance is split into two parts $C_p$ and $C_s$, one part with a significant series inductance and one with no series inductance. This split is a simplification that gives just one series resonant frequency (SRF) and one parallel resonant frequency (PRF). The inductance is associated with the fine conductors of the multi-layers and hence has an inductance higher than that for the resistor equivalent circuit of Fig. 7. The capacitance values $C_p$ and $C_s$ will be much bigger than the capacitance between the resistor terminals in Fig. 7. The consequence of large inductance and capacitance means that resonances will be at much lower frequencies. Values for an example equivalent circuit from Johanson Technology [9] for what would be a 10 pF capacitor at low frequencies, are $C_s$ = 4.8 pF, $C_p$ = 5.2 pF, $L_s$ =0.5 nH, $R_s$ = 0.11 Ω and $R_p$ = 78 kΩ. The current for an applied voltage of one is plotted in Fig. 9. Below 1 GHz the current rises as jωC. Above 1 GHz the effective capacitance starts to rise above 10 pF. The series resonance is at 3.25 GHz and at this point, the current would be 9 Amps (if 1 Volt could be maintained). Across the series resonance the phase across the circuit changes from -90º to +90º indicating that the capacitor is now an inductor. The parallel resonance occurs at 4.5 GHz. Near this frequency, losses are too high for the component to be useful. Sometimes a capacitor might be used between its series resonant frequency and its parallel resonant frequency as a DC blocking inductor.

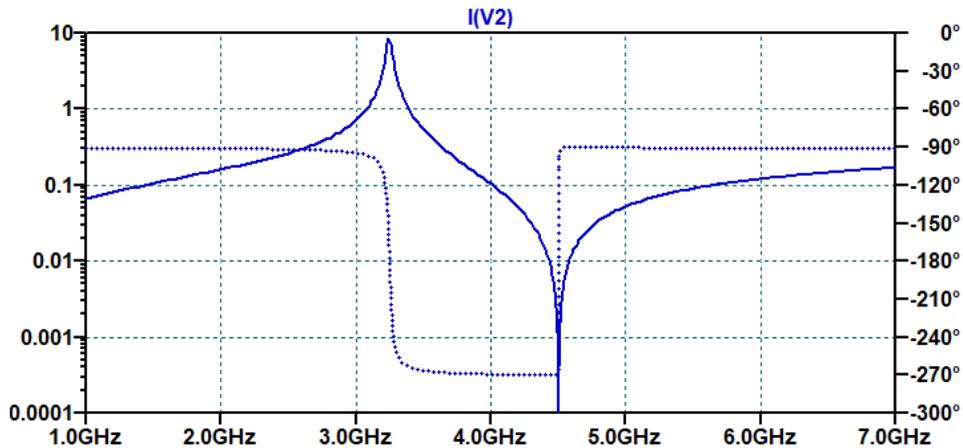

**Fig. 9:** Current through equivalent circuit of Fig. 8 representing a 10 pF capacitor as a function of frequency near its SRF and PRF.

Other basic RF components available as surface mount include inductors, filters and transformers. In Figure 6 (right) the white components on the two tracks to the far left and right of the board are Mini-circuits HFCN-103+ high pass filters to block intermediate frequencies leaking to the inputs on the two lower mixers. The footprint for this device requires side-earth connections across the GCPW gaps. The board has an Analog Devices AD8099 operational amplifier, whose Gain Band Width Product is 3.8 GHz. It is powered with +/-5 Volts and the supply lines have large surface mount series chokes and parallel electrolytic capacitors to reduce ripple. The amplifier also has other resistors and capacitors to set the gain and optimise performance at the chosen gain. It should be noted here that Analog Devices as well as being a supplier of a huge range of RF components with detailed application notes, they also provide volumes of excellent educational material for RF designers.

## 4    Printed structures

Signals from an accelerator system invariably require some processing. Where the frequency is too high for surface mount components, passive structures can be printed onto a PCB to act as inductors, small value capacitors and to undertake a few important functions, these include:
–    Stubs for impedance matching,
–    Filters (Low Pass, High Pass, Band Pass, Band Stop),
–    Splitters (Wilkinson),
–    Couplers (Directional, Branch Line, Hybrid Ring),
–    Resonators.

Four examples of printed components in microstrip from this list are shown in Fig. 10.

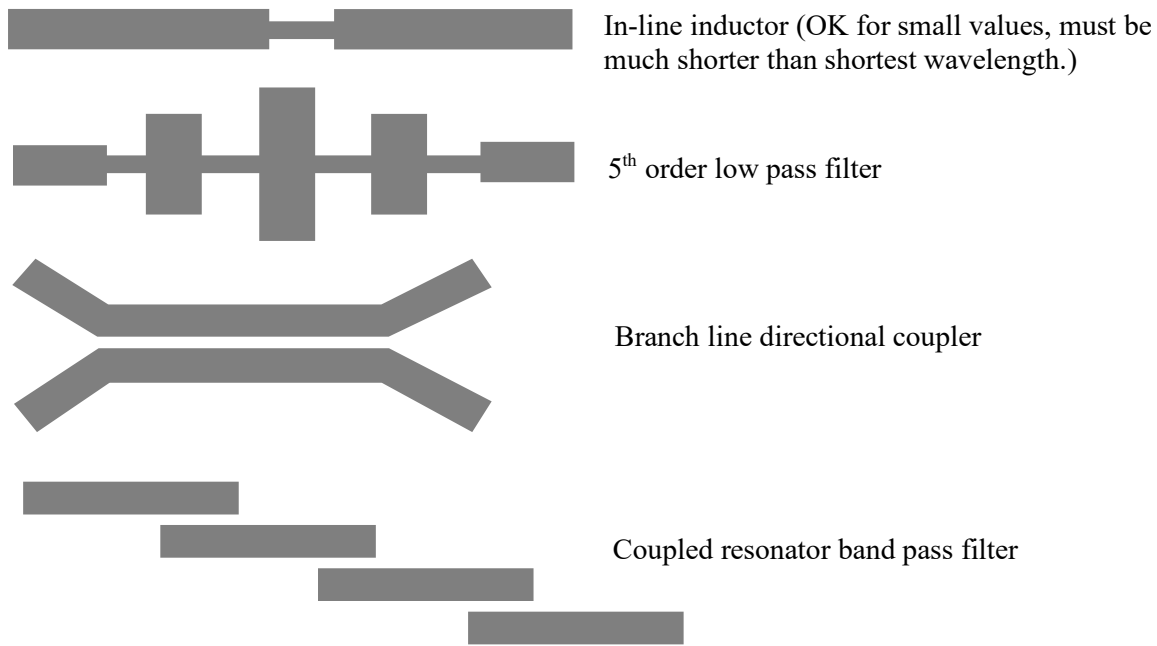

In-line inductor (OK for small values, must be much shorter than shortest wavelength.)

$5^{th}$ order low pass filter

Branch line directional coupler

Coupled resonator band pass filter

**Fig. 10:**  Printed microstrip components

Designing a structure to give the required S-Matrix to a high degree of accuracy requires considerable effort and it is for this reason that accelerator system engineers often choose to use connectorized components as shown in Fig. 11, where the performance is defined on a datasheet by the manufacturer.

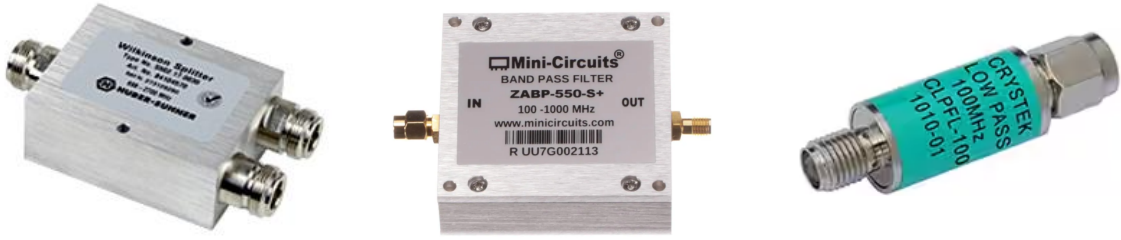

**Fig. 11:** Connectorized components

Dissecting these passive connectorized components one would see that often the chosen transmission line in use is suspended strip-line where microstrip tracks are mirrored and there is dielectric and an earth plane above and below the pair of mirrored tracks making the performance similar to a standard strip-line. The advantages of strip-line are lower losses hence higher Q structures and reduced cross talk between tracks.

Design equations for fabricating passive structures exist for all the preferred transmission lines including microstrip, co-planar waveguide and strip-line [10-13]. These equations are usually based on equivalent circuit representations. Precise design invariably needs EM simulation and testing for the substrate and transmission line of interest. Bespoke EM design tools include Keysight Pathwave Advanced Design System (ADS), Cadence AWR Microwave Studio, Altium Designer, Dassault CST Studio Suite, Siemens (PADS), ANSYS HFSS and Ansoft Designer [14-21].

### 4.1 Impedance discontinuity

In this section, the series inductor structure of Fig. 10 with two impedance discontinuities will be examined at a fixed frequency whose wavenumber is k. When the connecting section is short and the discontinuity is small then the figure also represents a suitable model for a connector as given with respect to Fig. 3. At a change of transmission line impedance, the voltage for the incoming and outgoing waves must balance and current is conserved across the junction. For a discontinuity from impedance $Z_1$ to impedance $Z_2$ let the amplitude of the incoming wave from $Z_1$ be $F_1$, the reflected amplitude be $R_1$ and the transmitted amplitude be $F_2$ also assume the outgoing transmission line is terminated so there is no reflection. Using Eq. (3) for the voltage at the junction $F_1 + R_1 = F_2$ and Eq. (4) to conserve current $(F_1 - R_1)/Z_1 = F_2/Z_2$. Solving gives the well-known equations,

$$R_1 = \left(\frac{Z_2 - Z_1}{Z_1 + Z_2}\right) F_1 \quad (9) \qquad F_2 = \left(\frac{2Z_2}{Z_1 + Z_2}\right) F_1 \quad (10)$$

Consider now both discontinuities as shown in Fig. 12 where the line to the right with impedance $Z_1$ is perfectly terminated so there is no reflection. In this case, there is now a reflection on the centre section as well as the left-hand section. Take the waves to be sinusoidal with wavenumber k. The forward wave $F_2$ that leaves the first discontinuity gains a phase of exp(-jka) when it arrives at the second discontinuity and the reflected wave $R_2$ also gains a phase of exp(-jka) when it arrives back at the first discontinuity where a is the length of the higher impedance section $Z_2$.

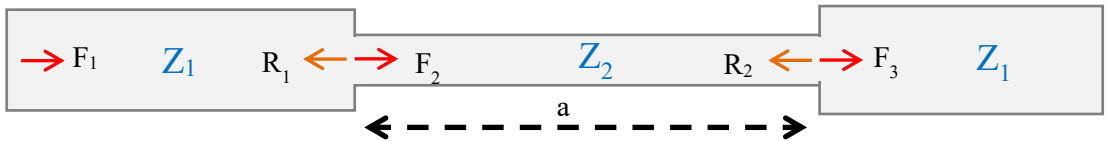

**Fig. 12:** In-line inductance

Applying Eqs. (9) - (10) at the second discontinuity gives,

$$F_3 = \left(\frac{2Z_1}{Z_1+Z_2}\right) F_2 e^{jka} \quad (11), \qquad R_2 = \left(\frac{Z_1-Z_2}{Z_1+Z_2}\right) F_2 e^{jka} \quad (12)$$

and at the first discontinuity gives,

$$F_2 = \left(\frac{2Z_2}{Z_1+Z_2}\right) F_1 + \left(\frac{Z_1-Z_2}{Z_1+Z_2}\right) R_2 e^{jka} \quad (13), \qquad R_1 = \left(\frac{Z_2-Z_1}{Z_1+Z_2}\right) F_1 + \left(\frac{2Z_1}{Z_1+Z_2}\right) R_2 e^{jka}. \quad (14)$$

Solving Eqs. (11) - (14) for $F_3$ in terms of $F_1$ gives,

$$F_3 = \frac{e^{jka}}{1 - A\left(e^{2jka} - 1\right)} F_1 \quad (15), \qquad \text{with } A = \frac{(Z_2/Z_1 - 1)^2}{4 Z_2/Z_1}. \quad (16)$$

Figure 13 plots the magnitude of $|F_3/F_1|$ given by Eq. (15) to show that the higher impedance track has a similar effect to an inductor. For a small length as compared to the wavelength (ka small), there is almost full transmission. For a set length with ka < π/2 then for increasing frequency (increasing k) transmission is reduced. Very high levels of attenuation is only possible if the impedance ratio $Z_2/Z_1$ is very large which make the track extraordinarily narrow hence this type of inductor is rarely used.

A small in-line inductor L on the transmission would give,

$$F_3 = \frac{F_1}{1 + j\omega L/Z_o}, \quad (17)$$

where $Z_o$ is the line impedance hence transmission always tends to zero at high frequencies.

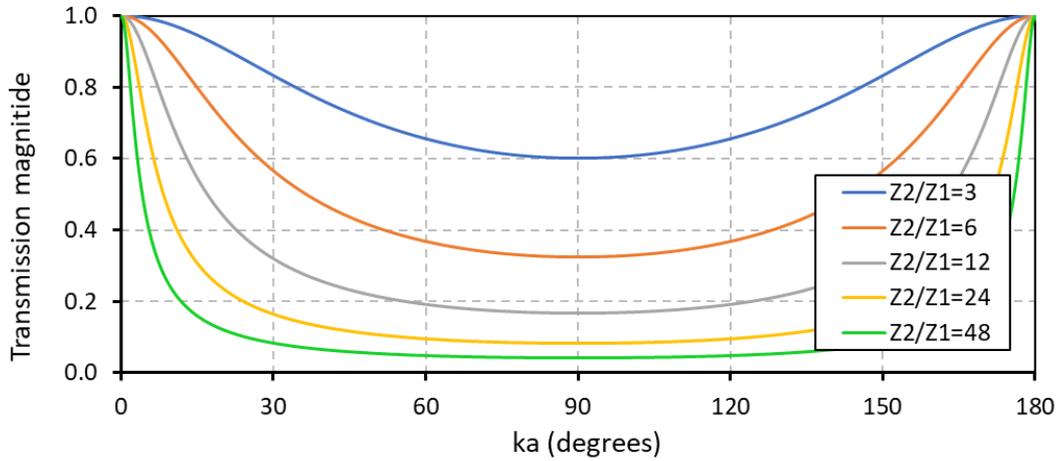

**Fig. 13:** Transmission of an in-line microstrip inductor

### 4.2 Transmission line stubs

A stub is a section of transmission line that is only connected to the signal line at one end, the other end is usually terminated with an open circuit or a short circuit. For most PCB transmission line technologies, including microstrip and grounded co-planar waveguide stubs can only be physically connected as a shunt between the signal and ground rather than as a series element.

If a transmission line stub with intrinsic impedance $Z_o$ is terminated with a load $Z_L$ then the impedance as a function of distance along the stub from the termination at z = 0 can be determined from Eq. (5). At the termination the ratio of the current to the voltage is determined by the load impedance hence Eq. (9) can be re-written as where $Z_1$ become the line impedance $Z_o$ and $Z_2$ become the load impedance $Z_L$. This gives,

$$R(Z_L + Z_o) = F(Z_L - Z_o). \tag{18}$$

Putting Eq. (18) in Eq. (5) gives

$$Z(z) = Z_o \frac{(Z_o + Z_L)\exp(jkz) + (Z_L - Z_o)\exp(-jkz)}{(Z_o + Z_L)\exp(jkz) - (Z_L - Z_o)\exp(-jkz)} = Z_o \frac{Z_L \cos(kz) + jZ_o \sin(kz)}{Z_o \cos(kz) + jZ_L \sin(kz)}. \tag{19}$$

The impedance of open and short-circuited stubs of length a with $ZL = \infty$ and $ZL = 0$ respectively become,

$$Z_{open} = -jZ_o \cot(ka) \tag{20} \quad \text{and} \quad Z_{short} = jZ_o \tan(ka). \tag{21}$$

For stubs with length a such that $ka < \pi/2$ the signs of Eqs. (20) and (21) imply that $Z_{open}$ acts as a capacitor and $Z_{short}$ acts as an inductor. When the length of the stub is chosen to be one-eighth of a wavelength at the frequency of interest so that $ka = \pi/4$ then,

$$Z_{open} = -jZ_o \quad \text{and} \quad Z_{short} = jZ_o.$$

Increasing the wavenumber k and hence the frequency above the frequency of interest for the one-eighth wavelength stub at the frequency of interest, then $Z_{open}$ decreases as reciprocal frequency near this point as expected for a capacitor and $Z_{short}$ increases in proportion to frequency near this point like an inductor.

## 5    Filters

### 5.1    Types

In communication systems filters are used to separate channels, which when closely spaced the filter needs a very sharp cut-off. For accelerator systems, the requirement is usually to remove well-separated, unwanted frequencies generated by non-linear components. When using a mixer or multiplier to generate new frequencies, more than one is always created.

Often for an accelerator system, one wants a smooth and almost flat response in the frequency range of interest hence a Butterworth filter might be one's first choice. Choosing between Butterworth, Elliptic, Chebyshev, Bessel and M-derived filters etc. is important if one is designing from scratch, however, given the huge number of commercially available filters, designing one's own filter is rarely necessary.

Filters are fabricated in a number of ways, principally they are lumped LC, LTCC (low temperature co-fired ceramic), ceramic, thin film, Surface Acoustic Wave (SAW), MMIC, PCB, tubular co-axial, saw, waveguide and cavity. Figure 14 gives images of a ceramic, a thin film and a tubular lumped element filter.

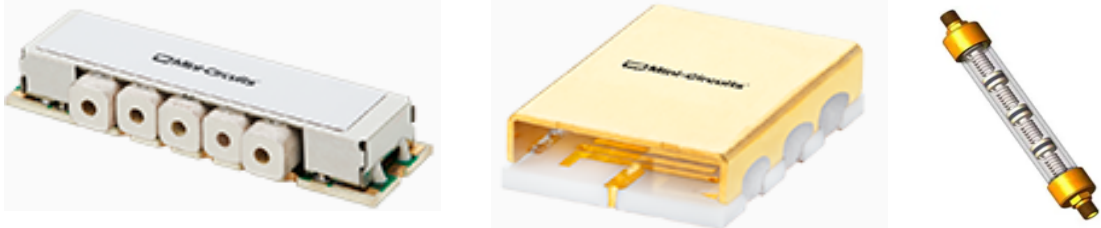

**Fig. 14:** Images of commercial Minicircuits ceramic and thin film filters (left and centre) and a Spectrum Control tubular lumped element filter (right).

Manufacturers of commercial filters will have product selection tools where only the final performance is important. If the design is being built on a PCB one might select an LTCC surface mount component as shown in Fig. 15. The component shown is a Minicircuits BFCN-3010+ with a price tag of ~5 Euro. It is designed to fit with co-planar waveguide and has 3 terminals, 2 on the line and one to earth. The figure also shows the PCB footprint. The earth passes under the component, consequently the gap for the co-planar waveguide is independent of the component design. The important selection criteria come from the filter response defined in the figure at seven marked frequencies and DC. Minicircuits supply LTCC components throughout the frequency range of about 400 MHz to 50 GHz [22]. It should be noted that thin film and MMIC filters are also available as surface mount and typically have very low reflection. For low frequencies < 400 MHz lumped LC filters are most commonly listed. The ceramic filters give the benefit of narrow passbands.

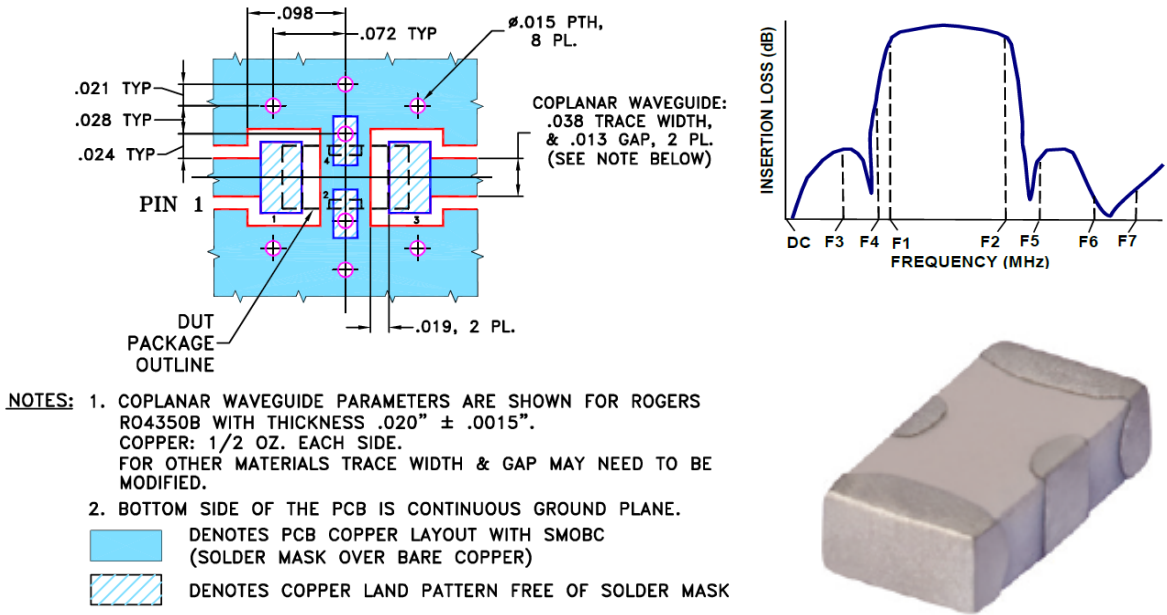

**Fig. 15:** Mini-circuits BFCN-3010+ LTCC surface mount bandpass filter.

## 5.2 Kuroda identities

Filters are usually designed starting from a lumped circuit of capacitors and inductors. The circuit can only be implemented with the ideal response if those capacitors and inductors are much smaller than the wavelength. As frequency increases and the wavelength reduces there becomes a point where a distributed analysis is required that takes account of the size and shape of the component. A transmission line stub is a distributed circuit with a well-defined response and can act as a capacitor or an inductor for a limited range of frequencies. The Richards transformation is used to map the response of a filter design realised with lumped elements to one realised with waveguide stubs. The responses are only similar close to the frequency of interest. Richards chose stubs whose length are one-eighth of the wavelength at the angular frequency of interest $\omega_o$ so that for other wave numbers k and associated frequencies, $\omega$ one gets,

$$ka = \tfrac{1}{4}\pi\omega/\omega_o .\tag{22}$$

For an inductor the Richards transformation is,

$$j\omega L \Rightarrow jZ_o \tan\left(\tfrac{1}{4}\pi\omega/\omega_o\right),\tag{23}$$

and for a capacitor the Richards transformation is,

$$1/j\omega C \Rightarrow -jZ_o \cot\left(\tfrac{1}{4}\pi\omega/\omega_o\right).\tag{24}$$

The Richards transformation maps the lumped element frequency response from the range $0 \le \omega < \infty$ to the range $0 \le \omega < 2\omega_o$. When using transmission line stubs, the filter response becomes periodic which unfortunately means there can be multiple pass bands when only one is desired. If the filter does not exclude all the unwanted frequencies it will need to be used with additional filters to block other ranges. Choosing the stub length to be one-eighth of the wavelength at the frequency of interest is not essential but it does give the most compact filters.

Where a lumped parameter filter is constructed as a ladder then it will have series capacitors and inductors as well as shunt capacitors and inductors. It is geometrically impossible to implement series stubs with microstrip or co-planar waveguide because the ground plane provides the return path. Kuroda found a way of replacing series inductors with shunt capacitors by inserting additional lengths of transmission line of a prescribed intrinsic impedance $Z_{UE}$ to achieve a specific component value. These additional lengths of transmission line are referred to as unit elements. The replacements are known as Kuroda's identities and are most easily derived using ABCD matrices.

For a two-port network, an ABCD matrix determines the voltage across and the current into one port of a network, as a function of the voltage across and the current out of the other port of the network. The ABCD matrices for networks composed of just one series or one shunt impedance Z are,

$$\begin{bmatrix} v_{in} \\ i_{in} \end{bmatrix} = \begin{bmatrix} 1 & Z \\ 0 & 1 \end{bmatrix} \begin{bmatrix} v_{out} \\ i_{out} \end{bmatrix} \quad \text{and} \quad \begin{bmatrix} v_{in} \\ i_{in} \end{bmatrix} = \begin{bmatrix} 1 & 0 \\ 1/Z & 1 \end{bmatrix} \begin{bmatrix} v_{out} \\ i_{out} \end{bmatrix} \quad \text{respectively.}$$

Ladder circuits can be analysed by multiplying ABCD matrices. For a transmission line of intrinsic impedance $Z_{UE}$ the ABCD matrix is easily determined by expressing the voltages and currents in terms of the forward and backward waves as was done to obtain Eq. (5). The result is Eq. (25) where the forward and backward phasors at the output are taken to be F and R. One requires that the forward wave loses phase $e^{jka}$ at the input and the reflected wave gains this phase, in this way the ABCD matrix can be deduced,

$$\begin{bmatrix} Fe^{jka} + Re^{-jka} \\ (Fe^{jka} - Re^{-jka})/Z_{UE} \end{bmatrix} = \frac{1}{2} \begin{bmatrix} e^{jka} + e^{-jka} & Z_{UE}(e^{jka} - e^{-jka}) \\ (e^{jka} - e^{-jka})/Z_{UE} & e^{jka} + e^{-jka} \end{bmatrix} \begin{bmatrix} F+R \\ (F-R)/Z_{UE} \end{bmatrix}. \tag{25}$$

Applying Eq. (22) so that the ABCD matrix is specific to the one-eighth wavelength stub at the frequency of interest, and simplifying with sines and cosines, the ABCD matrix for the unit element can be written,

$$[ABCD] = \cos\left(\tfrac{1}{4}\pi\omega/\omega_o\right) \begin{bmatrix} 1 & jZ_{UE}\tan\left(\tfrac{1}{4}\pi\omega/\omega_o\right) \\ j\tan\left(\tfrac{1}{4}\pi\omega/\omega_o\right)/Z_{UE} & 1 \end{bmatrix}. \tag{26}$$

The first two Kuroda identities are shown in Fig. 16 where

$$N = 1 + Z_{UE2}/Z_{UE1}, \tag{27}$$

and the unit matrix is any matrix of the form,

$$\begin{bmatrix} 1 & SZ_{UE} \\ S/Z_{UE} & 1 \end{bmatrix}. \tag{28}$$

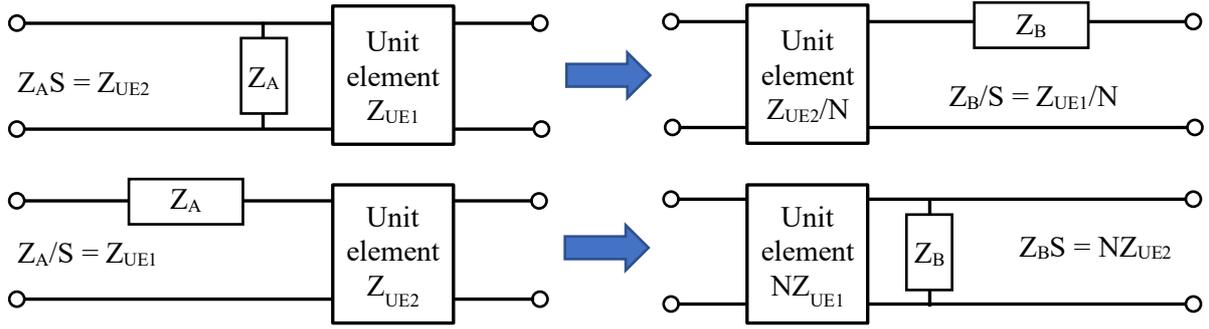

**Fig. 16:** Kuroda's first identity top and second identity bottom.

Impedances $Z_{UE} = Z_{UE1}$ or $Z_{UE2}$ are intrinsic impedances of transmission line sections and hence are real. Note that Eq. (28) does not quite map onto Eq. (26) as the cosine term is missing, the normalisation is now encompassed by the factor N given in Eq. (27).

The identities work for any functional form of S. If inductors are required as series components in the top right and bottom left images, then S must be complex and positive (i.e. S = j for instance). When S is complex and positive, then the shunt components in Fig. 16 must be capacitive as S multiplies $Z_A$ and $Z_B$ so would cancel a j in the denominator, for example, $1/j\omega C$ to give a real positive value for $Z_{UE}$.

If the capacitors and inductors are transmission line stubs then the Richards transformation will have been applied so that,

$$S = j\tan\left(\tfrac{1}{4}\pi\omega/\omega_o\right). \qquad (29)$$

In Figure 16 the lumped impedances $Z_A$ and $Z_B$ are the values calculated for the lumped circuit. As can be seen from Eqs. (26) and (28) the Unit element carries the frequency dependence with the function S.

The Kuroda identities are typically used to map a series inductance into a shunt capacitance so it can be replaced with microstrip or CPWG. The first identity maps a known shunt component with impedance $Z_A$ into series impedance $Z_B$. The shunt $Z_A$ can start on either side of the unit element, the identity then maps it to the other side, changing its value and also the impedance of the unit element. At the outset, one knows the value of $Z_A$ and $Z_{UE1}$. The value of $Z_{UE2}$ is determined from $Z_A$. The value of N is determined from $Z_{UE1}$ and $Z_{UE2}$. The value of $Z_B$ is determined from N and $Z_{UE1}$.

Kuroda's second identity maps a series impedance into a shunt impedance. At the outset, one knows the value of $Z_A$ and $Z_{UE2}$. The value of $Z_{UE1}$ is determined from $Z_A$. The value of N is determined from $Z_{UE1}$ and $Z_{UE2}$. The value of $Z_B$ is determined from N and $Z_{UE2}$.

Both identities are needed, and not just the second identity, because implementation on a PCB requires that all the shunt stubs are separated by unit elements. Geometrically it is impossible for finite-width shunt stubs to be adjacent. The example to follow shows this.

Kuroda's third and fourth identities (not given here) allow the values of series and shunt impedances to be transformed into more convenient values by adding a transformer.

The process of transforming a 5th-order low pass filter is shown in Fig. 17 with the top line giving the lumped circuit. Each subsequent line shows a transformation. The lumped circuit is drawn with series inductors and shunt capacitors as would be for a low pass filter. The transformations will work for any impedances hence they are written as $Z_1$, $Z_2$, $Z_3$, $Z_4$ and $Z_5$.

Often a filter design starts by using a table that applies for unit frequency, unit load impedance and unit generator impedance. The transformation to stubs and the application of the Kuroda identities can be made before any scaling to the frequency and line impedances of interest. Here impedances that could apply to the scaled or unscaled filter circuit are written.

The first step to produce line 2 in Fig. 17 is to place unit elements on either side of the filter with the line impedance that connects the generator and loads. Extending the transmission line on either side of the filter has no effect on the filter properties. Moving from line 2 to line 3 the first Kuroda identity has been applied. In this case, shunt capacitors have been converted to series inductors. This might seem curious as the challenge is to eliminate the inductors. Moving from line 3 to line 4 the input and output transmission lines are further extended. At this stage, every unit element is seen to have an adjacent series inductor. The second Kuroda identity in Fig. 16 is now applied to each of the four pairs so that all the series inductors become shunt capacitors in one step.

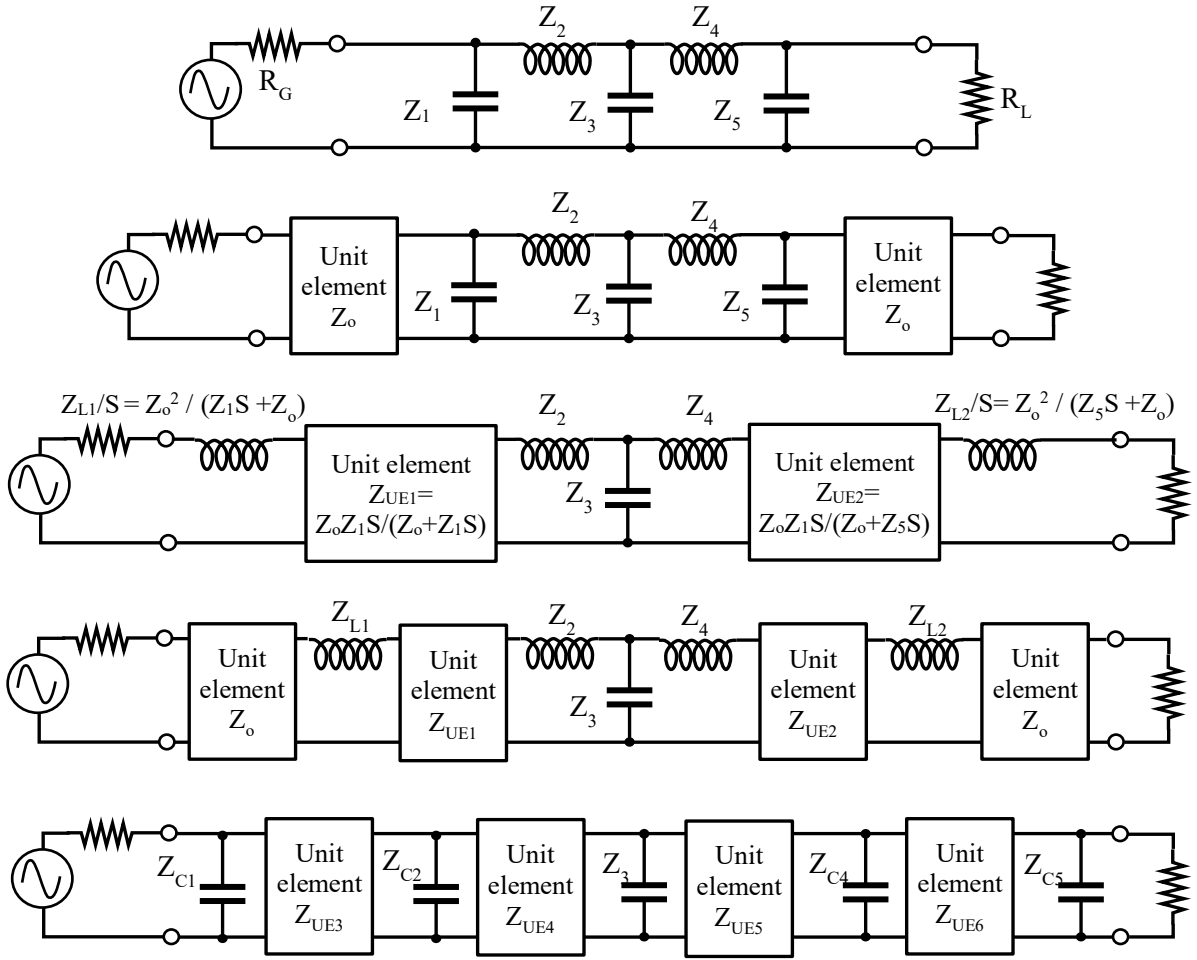

**Fig. 17:** Transforming a lumped circuit to one that can be implemented with transmission line stubs using the Kuroda identities.

In Figure 17 going from lines two to three, the application of the first Kuroda identity gives Eqs. (30) and (31) for determining component and unit element values,

$$Z_{L1}/S = Z_o^2/(Z_o + Z_1 S), \qquad Z_{UE1} = Z_o Z_1 S/(Z_o + Z_1 S), \qquad (30)$$

$$Z_{L2}/S = Z_o^2/(Z_o + Z_5 S), \qquad Z_{UE2} = Z_o Z_1 S/(Z_o + Z_5 S). \qquad (31)$$

In Figure 17 going from lines four to five, the application of the second Kuroda identity gives the Eqs. (32) - (35) for determining component and unit element values,

$$Z_{UE3} = Z_o + Z_{L1}/S, \qquad Z_{C1}S = Z_o + \frac{Z_o^2}{Z_{L1}/S}, \qquad (32)$$

$$Z_{UE6} = Z_o + Z_{L2}/S, \qquad Z_{C5}S = Z_o + \frac{Z_o^2}{Z_{L2}/S}, \qquad (33)$$

$$Z_{UE4} = Z_{UE1} + Z_2/S, \qquad Z_{C2}S = Z_{UE1} + \frac{Z_{UE1}^2}{Z_2/S}, \qquad (34)$$

$$Z_{UE5} = Z_{UE2} + Z_4/S, \qquad Z_{C4}S = Z_{UE2} + \frac{Z_{UE2}^2}{Z_4/S}. \qquad (35)$$

For the implementation of the filter in line 5 of Fig. 17 then the capacitors as drawn are open waveguide stubs, one-quarter of a wavelength in length and with intrinsic impedance $Z_{Cx}S$ where x runs from 1 to 4 as given by Eqs. (32)-(35). These stubs are now separated by quarter wavelength transmission lines giving a structure that can be printed.

Band-stop filters can be derived from low-pass filters by replacing series inductors with parallel LC resonators and shunt capacitors with series LC resonators. Increasing the stub length from one-eighth of a wavelength to one-quarter wavelength creates a parallel resonator for the open series stub and a series resonator for the shunt stub.

## 6  Wilkinson splitter

The PCB shown in Fig. 6 (right) has three Wilkinson splitters. The board is designed to measure relative phase by two separate methods, between two 12 GHz signals coming in from the top connectors in the figure. For independent methods, the signals must be split, and this is achieved with symmetrical Wilkinson splitters on each input. The splitters in Fig. 6 use ¾ rather than ¼ wavelength transformers. The S matrix for such a splitter when perfect is given in Eq. (36) which relates inputs and outputs,

$$\frac{j}{\sqrt{2}} \begin{bmatrix} 0 & 1 & 1 \\ 1 & 0 & 0 \\ 1 & 0 & 0 \end{bmatrix} \begin{bmatrix} F_{input} \\ R_{out1} \\ R_{out2} \end{bmatrix} = \frac{j}{\sqrt{2}} \begin{bmatrix} R_{out1} + R_{out2} \\ F_{input} \\ F_{input} \end{bmatrix}. \qquad (36)$$

The input $F_{input}$ is split into equal outputs $R_{out1}$ and $R_{out2}$ which go on to the mixers. One measurement is made by direct mixing of the two input signals to give a DC output proportional to the phase difference. Two other measurements are made by mixing input signals with a local oscillator to provide intermediate frequencies that can be sampled. The path length to the left-hand splitter is 90° longer than the path to the right-hand mixer putting synchronous signals at the mixer input into quadrature (*top centre in figure 6 right*). The mixer gives zero DC output for quadrature inputs of identical amplitude. When there is a synchronisation error, the DC voltage output can be used to adjust line length restoring synchronisation. For high sensitivity, the DC output voltage is amplified. When the signals are of unequal amplitude then a correction must be made. When the phase difference is large, then the measurement amplifier saturates. The other measurements made by sampling the two input signals separately at an intermediate frequency provide a 180° range for phase and also give amplitudes sufficiently accurate for correction to be made to the high-precision DC measurement.

As this is a course on the Basics of RF Electronics it is of interest to examine how the Wilkinson splitter works. Figure 18 gives the effective circuit. The arms of the splitter are quarter wave transformers with intrinsic impedance $Z_o\sqrt{2}$ where $Z_o$ is the impedance of the input and output lines. Ports $P_2$ and $P_3$ are connected with a resistor whose value is twice the intrinsic impedance. The routing of the quarter wave transformers is often chosen to minimise cross-coupling, to keep the resistor small and to minimise any reflection associated with the line curvature. The techniques applied here to analyse the circuit can be applied to other passive structures.

Where a circuit is linear then new solutions can be constructed by adding other solutions that are simpler to determine. The S matrix gives the transmission from one port to any other port. For analysis of inputs to ports $P_2$ and $P_3$, odd and even excitations of the two ports at the same time will be considered, subsequently allowing separate excitation of each to be determined.

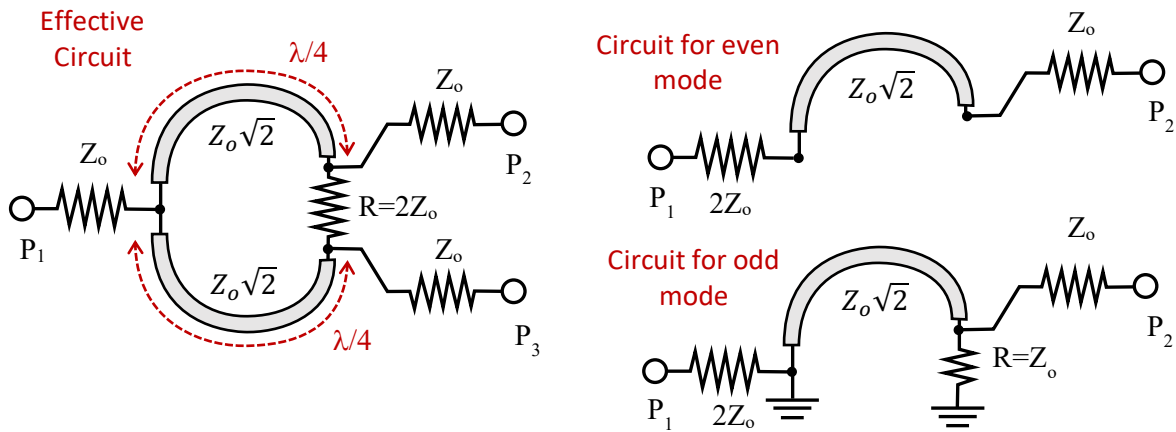

**Fig. 18:** Circuit for analysis of Wilkinson splitter, full circuit (*left*), even mode (*top right*), odd mode (*bottom right*).

Firstly, consider an input at port $P_1$. The splitter is symmetric hence the output must be identical at ports $P_2$ and $P_3$. This implies there will be no current in the resistor R. The input impedance can be considered alternatively as two parallel impedances of value $2Z_o$. Associating one parallel impedance with the top half and the other with the bottom half, then only the top half of the circuit needs to be drawn for analysis. This top half for the symmetric or even mode analysis is drawn top right in Fig. 18. In this figure it is seen that the input impedance of $2Z_o$ is now perfectly matched to the output impedance $Z_o$ with a quarter wave transformer. This solution gives no reflection at port $P_1$. As the S matrix is reciprocal then one also deduces that equal inputs at $P_2$ and $P_3$ add together at $P_1$ as an output. In order to construct an input to just $P_2$ or $P_3$ one can add an odd mode to the even mode,
i.e. $P_2 = 0.5 (P_2 + P_3) + 0.5 (P_2 - P_3)$. To complete the analysis, odd mode excitation must be determined. For odd mode excitation, the centre line must become an earth plane as shown in Fig. 18, bottom right. For the odd mode, the input $P_2$ sees a resistance of $Z_o$ to earth shunted with a shorted quarter wave transformer. From Eq. (21) a shorted quarter-wave transformer has infinite impedance hence the odd mode is perfectly matched with all the power being dissipated in the resistor. The S matrix of Eq. (36) follows. Fig. 19 shows the isolation obtained for the Wilkinson splitter on the test board in Fig. 6. The CPGW design parameters were modified after analysis of the test board so that maximum isolation occurred at 12 GHz. Transmission from ports $P_1$ to $P_2$ and $P_3$ has attenuation greater than 3 dB due to losses.

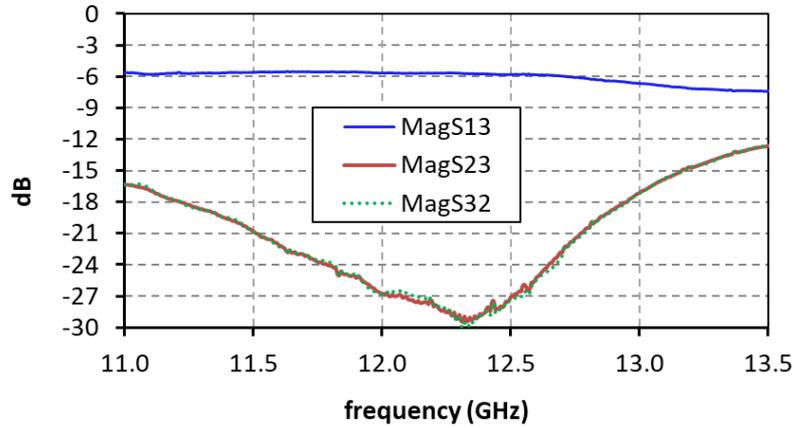

**Fig. 19:** Measurement of isolation and transmission for the Wilkinson Splitter shown on the test board in Fig. 6.

## 7 Mixers

Mixers perform time domain multiplication of signals. In the case of two signals being multiplied perfectly, then Eq. (37) which is identically true, shows that the output only contains the sum and difference frequencies of the two inputs,

$$2\cos(\omega_1 t+\theta_1)\cos(\omega_2 t+\theta_2)=\cos\{(\omega_1+\omega_2)t+\theta_1+\theta_2\}+\cos\{(\omega_1-\omega_2)t+\theta_1-\theta_2\} \quad (37)$$

The output phases are also sums and differences of the input phases. The output with the summed frequency is known as the up-converted signal and the output with the difference frequency is known as the down-converted signal.

If the input frequencies are identical then one output has double the input frequency coming from the up-converted term plus a DC offset that depends on the phase difference coming from the down-converted term. The offset is maximum when the phases are the same, and zero when they are 90º apart. A DC voltage level that determines phase difference can be extracted with a low-pass filter. For most accelerator systems accurate measurement of phase is critical. There will be a reference oscillator, and the phases of RF cavities must all be accurately set with respect to this reference. One issue therefore is to transport the reference to differing locations with knowledge of any phase shift with respect to the master reference. This problem is not considered here, the focus instead is on phase measurement.

Mixing to DC is rarely used as a method of measuring phase as errors arise from offsets in amplifiers and gains that depend on amplitude. Since the turn of the century, the most widely used method of measuring phase is by digitally sampling a waveform. Depending on the frequency of the RF signal of interest, this signal is usually down-converted before sampling. Down conversion is achieved by mixing the RF signal with a steady oscillator known as the local oscillator (LO). If the frequency of the local oscillator is relatively close to the frequency of the RF signal, then the up-converted signal is easily separated from the down-converted signal with a low pass filter. The down-converted signal is often referred to as the intermediate frequency (IF). For lower frequency down-converted signals, higher resolution can be achieved with digital sampling. For 16-bit sampling the AD9652 achieves 310 MSPS but for 12-bit sampling, the AD9207 achieves 6 GSPS [23].

Mixers are manufactured in huge quantities for communication and radar applications. An accelerator engineer would never need to design or fabricate a mixer. Two principal suppliers are Mini-circuits and Analog Devices. These companies provide mixer selection tables and detailed datasheets. For accelerator applications one normally desires good isolation between the RF port, the LO port and the IF port. Types of mixers that achieve good isolation are the double balanced mixers (DBM), triple-balanced mixers and Gilbert cells. The manufacturers' datasheets rarely give much information on the construction of their mixers or equivalent circuits. They do however provide extensive data relating to performance when used in communication systems.

The mixer most commonly selected for use in accelerators is probably the double-balanced mixer. Balanced implies they have a centre-tapped transformer on an input which provides a high degree of isolation. For high-frequency operation, the DBM invariably uses a ring of Schottky diodes to multiply the input signals. Schottky diodes have their junctions formed with a metal-to-semiconductor interface. This reduces the "turn-on voltage" well below that of a silicon PN junction and enables operation at much higher frequencies.

Figure 20 gives an LT Spice model of a fictitious DBM chosen so that output can be generated for a wide range of inputs. When a DBM is used within a communication system, RF1 is the LO and is typically excited at a level such that diodes D1 and D2 are fully conducting for the positive half cycle (*D3 and D4 not conducting*) and diodes D3 and D4 are fully conducting for the negative half cycle (*D1 and D2 not conducting*). This mode of operation is illustrated in Fig. 21, where the local oscillator current takes the black route and that part of the red route that competes a circuit back to the black route.

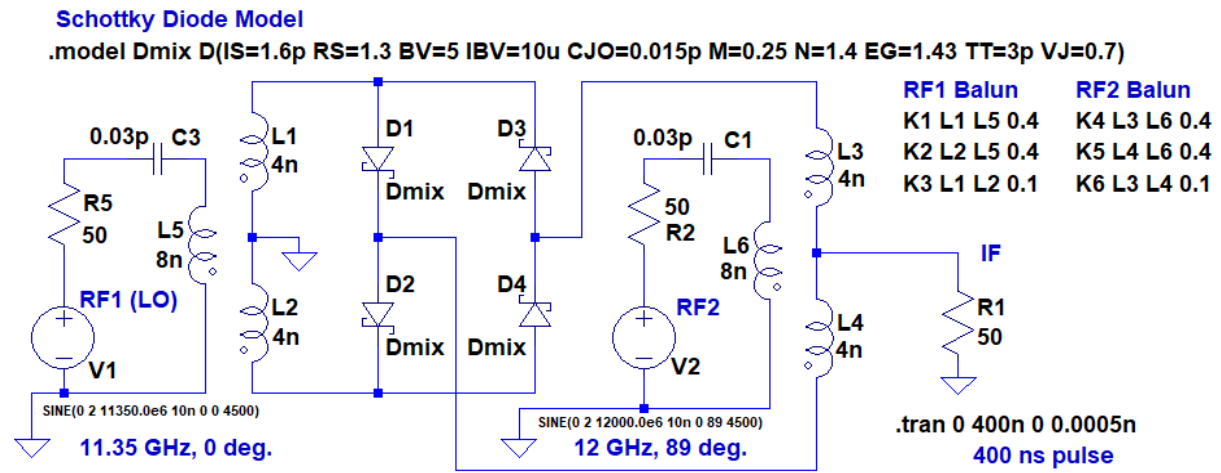

**Fig. 20:** LT Spice model for a double-balanced mixer.

The input RF2 in Fig. 20 will be a signal from a communication system or a waveform to be measured and is usually set at a level that is somewhat less than the LO level and insufficient to make the diodes conduct significantly by themselves. (*Note Fig. 20 shows identical levels of 2V for the RF and LO inputs*.) As the LO signal keeps one or the other of the diode pairs turned on for virtually the whole time, then the RF signal can flow through one of these pairs even though its peak voltage level is too small to turn the diodes on by itself. The polarity of the LO determines the route through the diode ring that the RF signal takes between the earthed centre tap on the LO balun and the IF output R1 (*Fig. 20*). The route through the RF balun (*on the right in Fig. 20 and at the bottom for Fig. 21*) determines the sign of the RF output.

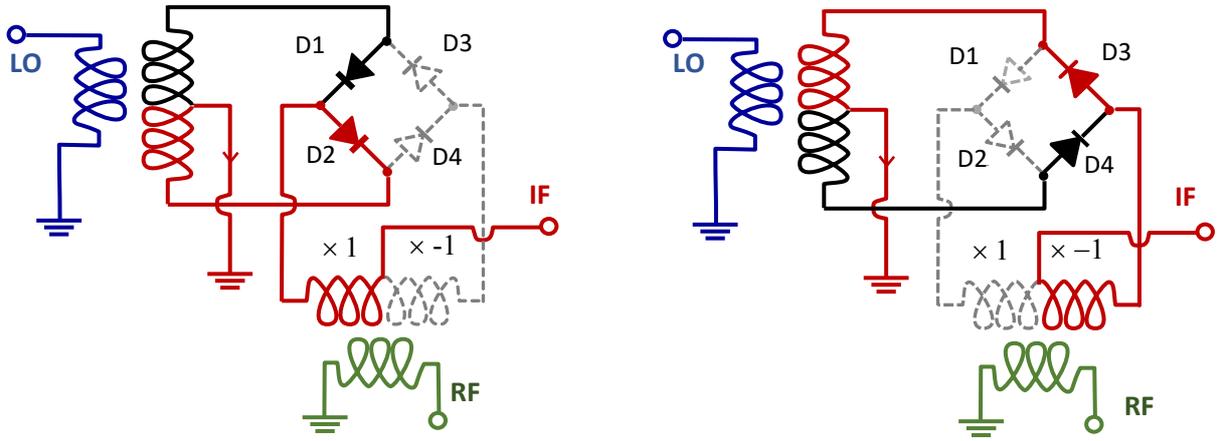

**Fig. 21:** Operation of the double-balanced mixer when the LO is at a level that makes diodes D1 and D2 fully conducting when positive (left) and diodes D3 and D4 fully conducting when negative and the RF is at a low level, insufficient to make any of the diodes conducting.

In Figure 21, for both left and right drawings the RF signal takes the red and black route. When the RF signal is much smaller than the LO signal then half takes the red route and half takes the black route by increasing or decreasing the LO current. As the two halves of the RF signal have opposite sense in the centre tapped balun then almost complete isolation to the LO input is achieved. The red and black route for the left-hand diagram goes through the RF balun leading to the IF port with the opposite sense to the red and black route on the right-hand diagram. This means that the RF signal gets multiplied by plus one for the positive half cycle of the LO and by minus one for the negative half cycle of the LO. Effectively the RF signal is multiplied with a square wave at the frequency of the LO. As will be seen from the algebra, this multiplication provides the mixing action and also the isolation between the RF input and the IF output. Eq. (38) multiplies the RF by a square wave given as a series,

$$V_{IF} = \frac{4}{\pi} V_{RF} \cos(\omega_{RF} t) \left\{ \cos(\omega_{LO} t) - \frac{1}{3}\cos(3\omega_{LO} t) + \frac{1}{5}\cos(5\omega_{LO} t) + \cdots \right\} \quad (38)$$

which can be expressed as

$$V_{IF} = \frac{2}{\pi} V_{RF} \left\{ \cos(\omega_{LO} - \omega_{RF})t + \cos(\omega_{LO} + \omega_{RF})t - \frac{1}{3}\cos(3\omega_{LO} - \omega_{RF})t - \frac{1}{3}\cos(3\omega_{LO} + \omega_{RF})t + \cdots \right\}$$

(39)

The first term is the down-converted frequency. If the frequency of the LO signal is close to the RF signal, then the second and third terms are close to double this frequency. When the down-converted frequency is very low, there is no output near the LO frequency or any odd order of the LO frequency.

It is of interest to see the outcome when the RF signal has the same amplitude as the LO, and both are sufficient to turn on the diodes. This has to be done by simulation. The RF frequency has been chosen as 12 GHz and the LO frequency as 11.35 GHz so that the output frequencies are sufficiently well-spaced to see the spectrum clearly. The FFT of the output is plotted in Fig. 22 and part of the associated waveform is plotted in Fig. 23.

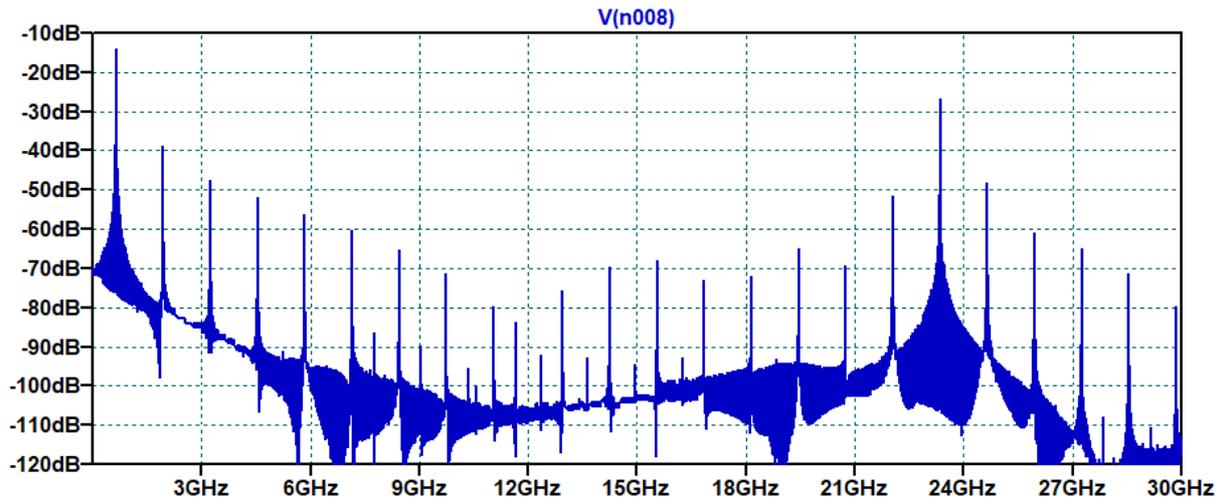

**Fig. 22:** FFT output of the circuit in Fig. 20 when the double-balanced mixer has equal amplitude inputs.

The down-converted frequency is at 650 MHz as expected but then there are additional but much smaller spectral peaks at 3, 5, 7, 9 etc times 650 MHz. These peaks could be highly undesirable for a communication system but would be easily removed with a filter for an accelerator system.

When mixing to DC (both RF and LO at 12 GHz) the first spectral peak is at 24 GHz and after low pass filtering the DC output voltage varies linearly with phase from zero at 90º to about 30º. This result requires the amplitudes to be identical. When measuring phase by mixing to DC the amplitude must be accurately measured and corrections applied.

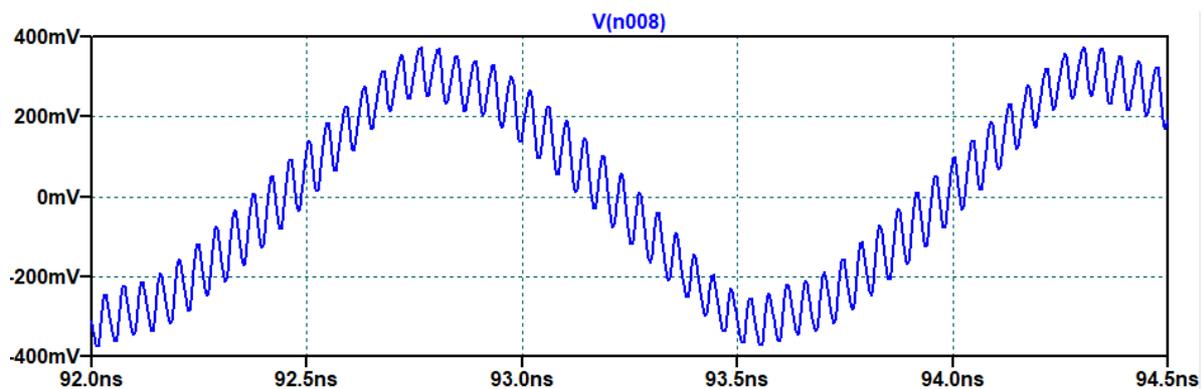

**Fig. 23:** Part of the waveform output of the circuit in Fig. 20 when the double-balanced mixer has equal amplitude inputs.

For the phase measurement board of Fig. 6, the lower two mixers used a LO frequency of 11.948 GHz, the RF was 11.994 GHz to give an IF of 46 MHz. The IF was sampled asynchronously at 120 MSPS.

Features for the SIM-24MH+ mixers used are:

- RF/LO frequency range        7.3 GHz to 20 GHz
- IF frequency range             DC to 7.5 GHz
- LO Power                      (+10 dBm to +16dBm) recommended =+13dBm
- Conversion loss               (5.84 dB at 12 GHz for +13 dBm LO)
- IP3                              (16 dBm at 12 GHz).

Figure 24 gives the LO to RF isolation (*left*) and LO to IF isolation (*right*) as a function of frequency for differing LO oscillator power levels.

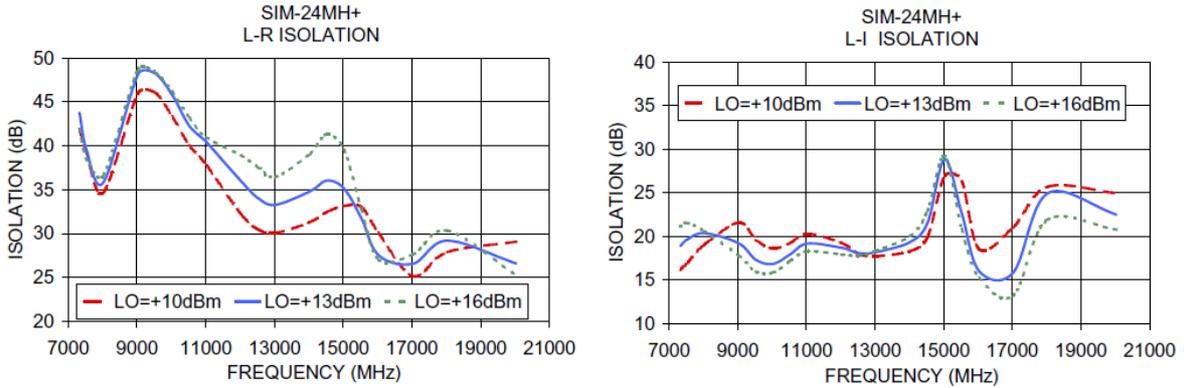

**Fig. 24:** Isolation for SIM-24MH+

For an accelerator system, LO power that finds itself as an outgoing wave on the RF input may not have any detrimental effect provided the connectors are reflectionless. This power should get completely absorbed in the load of the directional coupler that samples RF to the cavity. Note that couplings to high-power RF cavities for the purpose of measurement are typically -70 dB to -100 dB. RF that finds itself as an outward wave on the LO oscillator input needs to be absorbed somewhere rather than being reflected back. The simplest solution is to amplify the LO and then attenuate it.

Feed through of the LO or RF to the IF is easily removed by low pass resistive (RC) filtering when the mixer is down-converting. IF on the LO and RF line can be removed by high-pass resistive filtering.

Figure 25 gives the voltage standing wave ratio for the LO and RF inputs. An important observation is that the mixer is never matched. Reflected power is determined from the VSWR as,

$$P_R = \left(\frac{VSWR - 1}{VSWR + 1}\right)^2,$$

so, for a VSWR of 3 for instance then 25% of the power is reflected. Matching at the frequency of interest will change the phase that is to be measured hence is probably not useful. Some reflection of the LO may not be an issue if it never changes.

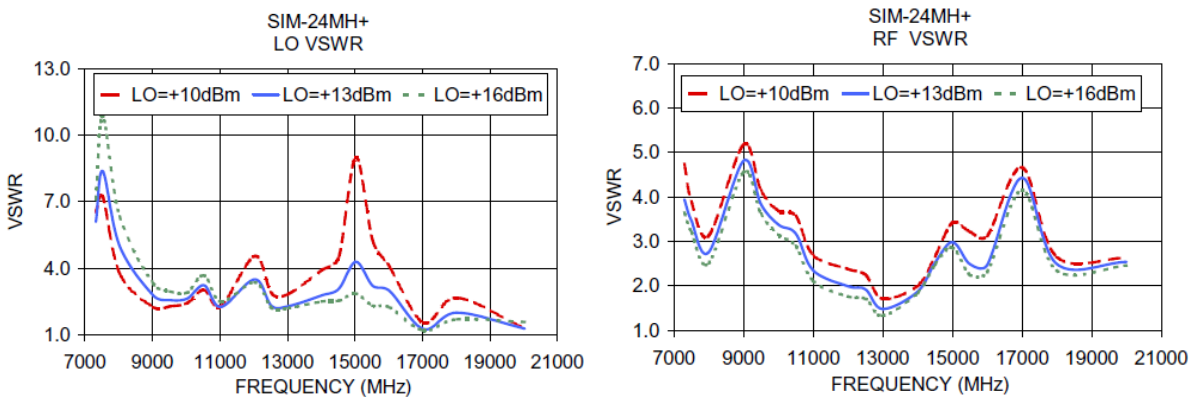

**Fig. 25:** VSWR for SIM-24MH+

# 8 I&Q modulation

In the expression I&Q (*or simply IQ*), I stands for that part of the signal that is in phase with the reference oscillator and Q stands for that part of the signal that is 90º retarded from the reference signal. In order to correct the phase of a cavity subject to perturbations from the beam and other influences, the phase and amplitude of the RF drive must be varied. One method to do this is using electronic phase shifters and variable attenuators on the signal from the master oscillator before it is inputted to the high-power amplifier train. The operating mode of an accelerating cavity is governed by a second-order differential equation. Whilst controlling separately on amplitude and phase works in general, it is not optimal in terms of achieving the best possible tracking of the set point.

If the time-varying cavity voltage V is decomposed with a steady frequency equal to the nominal operational frequency $\omega$ together with in-phase and quadrature, slowly varying functions of time $A_i(t)$ and $A_q(t)$ as given by Eq. (40),

$$V(t) = \{A_i(t) + jA_q(t)\}\exp(-j\omega t) \tag{40}$$

then for a high Q cavity, to an excellent approximation the functions $A_i(t)$ and $A_q(t)$ obey simultaneous first-order differential equations and hence the cavity model becomes a dynamical system. The drive for these two equations is simply the in-phase and quadrature components of the forward power delivered to the cavity. Proportional-integral (PI) controllers acting separately on the in-phase and quadrature components of the drive and dependent on errors in the in-phase and quadrature components of the cavity voltage is optimal.

In order to implement this controller, I and Q must be measured and then an IQ modulator applied to the drive. Figure 26 shows the internal structure of an IQ modulator. If the inputs and outputs are reversed, then functionally the modulator becomes an IQ detector.

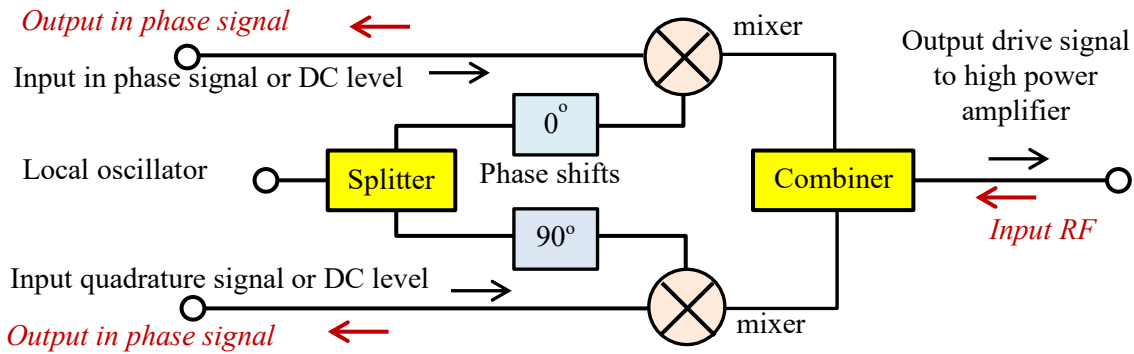

**Fig. 26:** IQ modulator (black arrows) or IQ detector (red arrows).

The detector is identified by the red arrows in Fig. 26. The preferred method of measuring I and Q for accelerator systems is by digitally sampling the waveform rather than using an IQ detector. The reason for this becomes apparent from the datasheets of real devices. Figure 27 gives the internal structure of the AD8346.

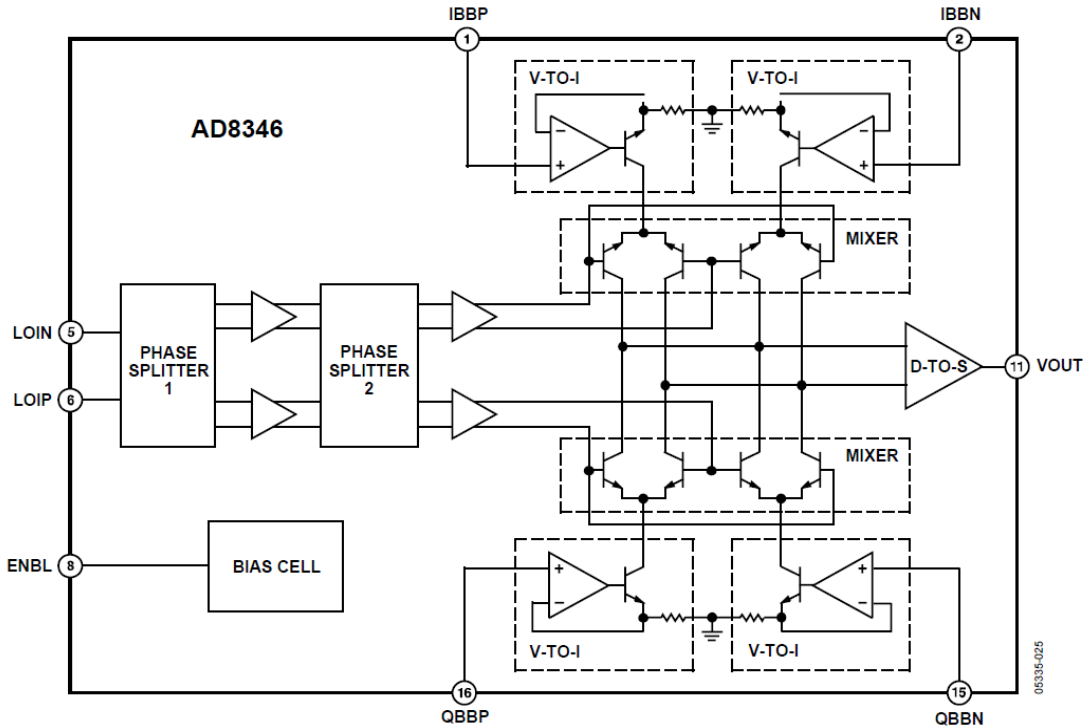

**Fig. 27:** AD8346 IQ modulator functional diagram.

It can be seen that this IQ modulator uses Gilbert Cells to perform the mixing operation. The in-phase, quadrature and LO inputs at the top, bottom and side respectively are differential inputs. If the in-phase and quadrature signal come from a DAC then the chip is likely to have differential outputs. A differential input for the LO may need a transformer.

Features for the AD8346 IQ modulator are:
- RF/LO frequency range           0.8 GHz to 2.5 GHz
- IQ modulation bandwidth         DC to 70 MHz
- Output Power                    -10 dBm (typical)
- LO power                        -12 dBm to – 6 dBm
- IQ Input impedance              12 kΩ
- Quadrature phase error           1-degree rms @ 1.9 GHz
- I/Q Amplitude balance           0.2 dB @ 1.9 GHz
- Sideband suppression            -36 dBc @ 1.9 GHz
- Noise floor                     -147 dBm/Hz.

A feature of immediate concern is the quadrature phase error of $1°$. Accelerator systems typically need phase control at the level of 5 to 100 milli-degrees rms. This is the reason why one might not choose an IQ detector chip to determine cavity phase in preference to digitally sampling. As a modulator, this quadrature phase error is far less problematic. Whilst the cavity must be controlled to milli-degrees because it has a high Q factor, a significantly larger phase shift in the drive is required to start making a milli-degree correction in the cavity. The correction is controlled independently for the I and Q components and the controller finds a drive level that gives the correct cavity phase and amplitude independently of any phase or amplitude imbalance in the modulator.

## 9 Oscillators

### 9.1 Application

This section covers oscillator basics but firstly it is of interest to put context on the use of oscillators. Figure 28 shows the frequencies and frequency conversions that are utilised for the plasma wakefield experiment AWAKE, where a proton beam is extracted from the CERN's SPS and injected into a plasma to accelerate bunches of electrons from an electron Linac. Experiments like AWAKE need multiple frequencies to be generated all of which must be synchronised at different levels of timing error.

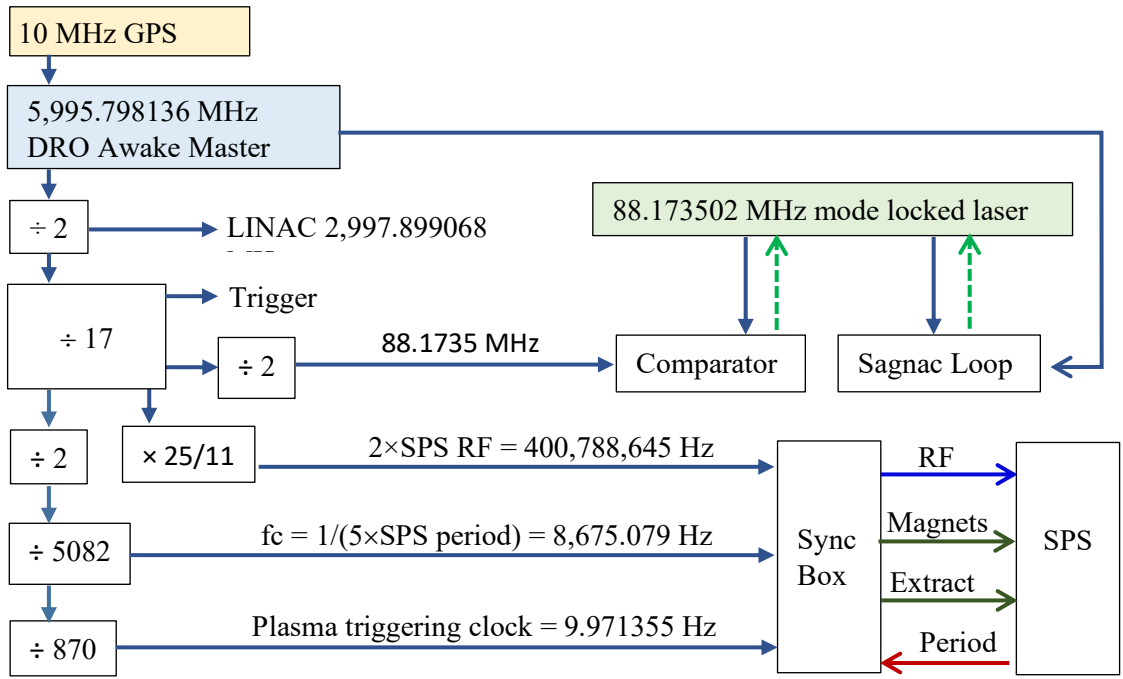

**Fig. 28**: AWAKE frequencies

The master oscillator is often an atomic clock that has excellent long-term stability and can be distributed around the experiment. Many atomic clocks contain a quartz crystal that is phase-locked to an atomic transition frequency. Master oscillators typically use a low frequency. The fundamental mode of an AT-cut quartz crystal has a practical frequency range of 600 kHz to 30 MHz.

The frequency stability of commercial, crystal stabilised, oscillators is between one part in $10^{-6}$ and one part in $10^{-7}$. The latter figure is about 1 second in 4 months or 1 minute in 20 years. If however, for example, one wants to time separate events occurring every 20 milliseconds (50 Hz) to an accuracy of 50 femtoseconds, a stability of $2.5 \times 10^{-12}$ is needed. Precision quartz oscillators held at constant temperature and protected from environmental disturbances, have fractional stabilities in the range $10^{-10}$ to $10^{-12}$. Mode-locked lasers can have significantly better fractional stability than $10^{-12}$. In Figure 28 the dielectric resonator oscillator (DRO) which is phase-locked to the 10 MHz GPS signal, is used to generate all the other necessary frequencies by integer division and multiplication. The mode-locked laser has better stability than the DRO and is used to time the laser pulses. It can also measure the stability of the DRO reporting on short-term phase errors.

In Figure 28 all the division operations use digital methods and hence produce square waves. Digital division is typically performed with D flip-flops. A wide range of commercial chips are available for division. Care must be taken to choose chips with low jitter. Note that the first divide by 2 on the Linac frequency could have been performed with a mixer and filter to produce a sine wave. This is not necessary as the amplifier train is very narrow band and a square wave input is transformed into a sine wave.

**9.2   Oscillator basics**

Reduced to its simplest format, an oscillator consists of an amplifier and a filter operating with positive feedback. Two variations on the format are shown in Fig. 29. The left-hand figure has a bandpass filter on the feedback path, hence one amplifies "more of the same" to get oscillation. Importantly, for amplification to occur, the phase gain around the whole loop must be 360° otherwise it is not "more of the same". This is the Barkhausen condition. The right-hand picture in Fig. 29 shows the essential condition for oscillation with a phase delay on the feedback path. A bandpass filter always gives a phase delay. The oscillator does not oscillate at the lowest impedance of the bandpass filter but rather at a frequency nearby, where the loop phase is 360°. If for instance, the amplifier is an inverting amplifier with a small time delay, then the amplifier might provide 185° in which case, the frequency adjusts itself until the phase delay in the bandpass filter is 175°. It should be noted that low pass, high pass and band stop filters all give phase delays and so can also be used to make oscillators.

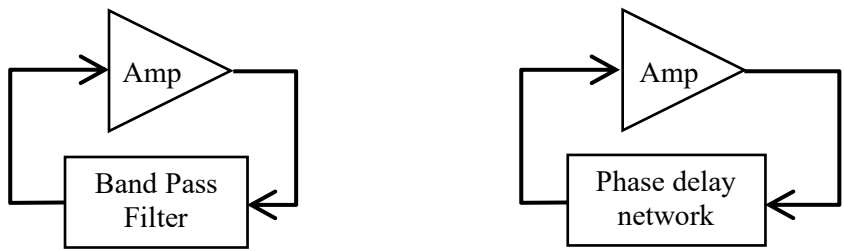

**Fig. 29:** Models for oscillator operation

Whilst an oscillator can be made with any phase delay network, for high stability operation one chooses a very high Q factor Band Pass Filter, which will typically be a resonant crystal, microwave cavity or ceramic block. The Q factor is defined as

$$Q_F = \frac{f}{\delta f}, \tag{41}$$

where f is the centre frequency and $\delta f$ is the 3dB bandwidth. When $Q_F$ is very large then the bandwidth is very small and hence the phase shift across the filter changes very rapidly in a small range. This means that the oscillator stabilises itself in this very small range. Crystals and resonators can be modelled with inductors (L), capacitances (C) and resistors. It is of interest therefore to consider LC lumped circuit resonators and phase delay networks first. Amplification might be provided by a three-terminal active device, for instance, a Bipolar Transistor or an FET. When driven at a particular frequency, series or parallel combinations of inductors and capacitances can be represented by a single impedance. The impedance can be taken as positive or negative as the oscillator is never perfectly on the resonant frequency of individual groups of components. With just three terminals the RF characteristics can be modelled with just three impedances as shown in Fig. 30. The arrow from the current generator is drawn upwards so the current feedback I through $Z_1$ is positive.

For oscillation at least one impedance must be capacitive and one must be inductive. The many various oscillator configurations such as Butler, Clapp, Colpitts, Hartley, Meacham, Miller, Seiler, and Pierce all have this underlying RF topology. Differences arise depending on the earth connection, where the output is taken and the signs of the impedances. For high Q operation, the impedances must be almost lossless. Pierce, Colpitts and Clap oscillators have two capacitors and one inductor, they differ in where the earth is positioned. Earth for the Pierce Oscillator is on the emitter or source. Earth for the Colpitts oscillator is on the base or gate. Earth for the Clapp is on the collector or drain.

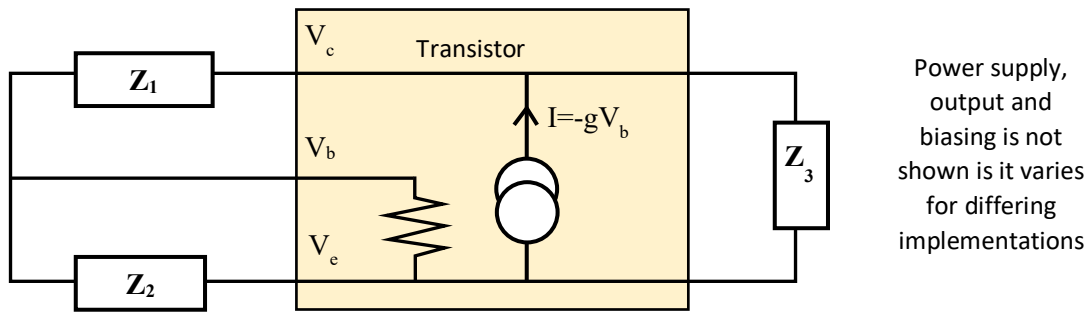

**Fig. 30:** RF model of single transistor oscillator.

Figure 31 shows a practical circuit for the Pierce oscillator using a bipolar junction transistor. The power rail that feeds the collector shares the feedback path. It also supplies the biasing resistors so that the transistor conducts. The stabilisation resistance gives negative feedback to limit DC flow from the collector to the base. The stabilisation resistor can also be a path for the RF but there is an alternative path for the RF through the linearisation resistor.

Comparing Fig. 31 with Fig. 30 it can be seen that $L_1$ is $Z_1$ as it connects the base to the collector. In series is the amplitude control resistor to limit feedback and a coupling capacitor with a very small impedance so it blocks DC from the biasing circuit. The capacitor $C_2$ is the impedance $Z_2$ as it connects the base to the emitter. It has a series resistance through the linearization resistor and the stabilisation resistor. The capacitor $C_3$ is the impedance $Z_3$ as it connects the collector to the emitter.

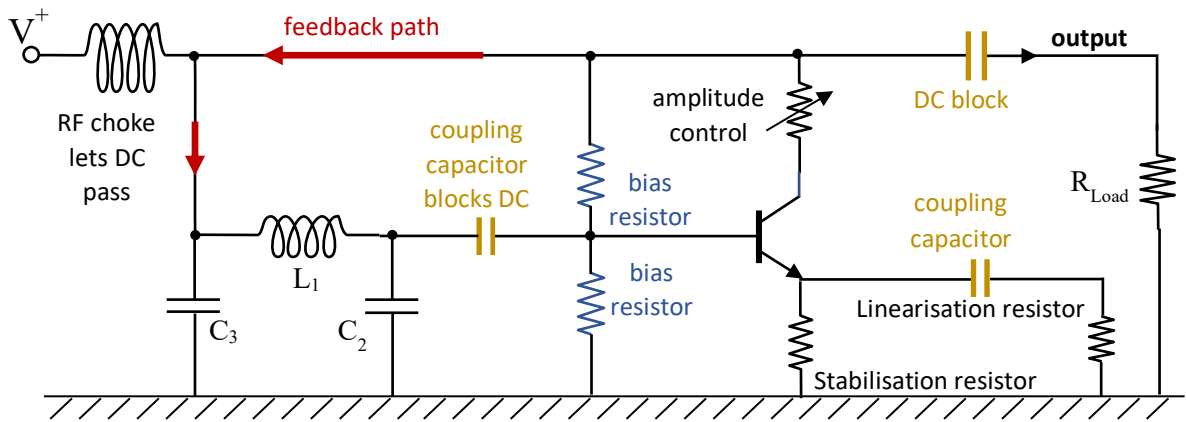

**Fig. 31:** Practical circuit for an LC Pierce oscillator.

When the principal impedances $L_1$ $C_2$ and $C_3$ are drawn as a ladder, then the filter type is apparent. In this case, the filter is a low-pass filter. Very low frequencies are not delayed, so arrive at the base 180° degrees out of phase. The corner frequency of the network is given as,

$$\omega_c = \sqrt{\frac{(C_1 + C_2)}{LC_1C_2}}$$

hence at this frequency, the phase shift is 180° bringing the circuit close to resonance. For this circuit, the frequency of oscillation is slightly below the corner frequency and depends on resistive loading. At microwave frequencies, transistor oscillators are designed from their S parameters. The frequency of the oscillation is most easily adjusted by changing the values of the capacitors $C_3$ or $C_2$.

## 9.3 Voltage controlled oscillators

A voltage-controlled oscillator (VCO) is an oscillator whose frequency can be changed by the application of a voltage. A reverse-biased PN junction diode does not conduct DC, it has two charged layers in close proximity separated by a depletion layer. In reverse bias, the diode becomes a capacitor. The width of the depletion layer grows with increasing voltage hence it is a variable capacitor. In Figure 31 a VCO might be constructed by replacing $C_2$ or $C_3$ with a varactor diode. A huge selection of VCOs are available from MW component suppliers. Mini-Circuits sells over 800 models covering the frequency range from 12.5 MHz to 8 GHz. For broadband operation below 2 GHz, the technology is likely to be LC-Monolithic Microwave Integrated Circuits (MMIC). Their selection table has headings as given in Table 2. Three products have been selected to illustrate typical performance.

Table 2: Part I - Selected Mini-circuits VCOs

| Model | Low Freq (MHz) | High Freq (MHz) | Power Output dBm | Tuning voltage range | Phase Noise (dBc/Hz) offset 1kHz | Phase Noise (dBc/Hz) offset 10kHz | Phase Noise (dBc/Hz) offset 100kHz | Phase Noise (dBc/Hz) offset 1MHz |
|---|---|---|---|---|---|---|---|---|
| ROS480+ | 386 | 480 | 9.5 | 3-16 | -88 | -115 | -137 | -158 |
| ROS1700 | 770 | 1700 | 9 | 1-24 | -72 | -99 | -120 | -141 |
| ROS3050 | 2635 | 3050 | 6.5 | 0.5-16 | -77 | -104 | -125 | -145 |

Table 2: Part II - Selected Mini-circuits VCOs

| Model | Pulling (MHz) pk-pk @12dB reflect | Pushing (MHz/V) | Tuning sensitivity (MHz/V) | Harmonics (dBc) typical | Harmonics (dBc) max | 3dB control bandwidth | DC Voltage (V) | DC (mA) |
|---|---|---|---|---|---|---|---|---|
| ROS480+ | 0.25 | 0.45 | 8-12 | -20 | -13 | 30 | 12 | 31 |
| ROS1700W | 7 | 0.7 | 26-65 | -25 | -9 | 50 | 12 | 35 |
| ROS3050C | 4.5 | 0.2 | 35-47 | -22 | -12 | 100 | 8 | 40 |

Important parameters on performance are the phase noise figures in columns 6 to 9 for part 1 of the table. Phase noise tells one about irregularity in the frequency. This topic is covered later. Inspection of the whole product table indicates that phase noise at fixed offsets increases with frequency and also with the bandwidth being provided. For accelerator applications, one is normally looking for narrow bandwidth VCOs.

In column 2 of table part 2, the pulling figure is given. An oscillator changes its frequency significantly when reflected power is not in phase or antiphase to the oscillation. If the load that the oscillator drives is unchanging, then pulling is not an issue after the correct frequency has been set. In column 3, part 2 of the table, the pushing figure is given. If the drive voltage changes then the frequency changes. If the maximum acceptable frequency departure is, say, 100 Hz, then for 0.2 MHz/V the power supply stability needs to be better than 0.5 mV. It is important to minimise ripple on all VCO power supplies. VCOs can have significant harmonics and so should be used with low-pass filters.

## 9.4 Crystal oscillators

Quartz crystals have piezoelectric properties and hence their electrical resonance is coupled to a mechanical resonance. Their fundamental mode can be accurately modelled with a series LCR circuit and a small parallel capacitance coming from the mounting as shown in Fig. 32. The capacitance $C_1$ is typically in the range $10^{-3}$ pF to $10^{-1}$ pF and is associated with piezoelectric charge movement as a function of voltage. The inductance depends on crystal mass and can be as large as a few Henrys for a large thick crystal down to a few milli-Henrys for a small crystal. Values of $R_1$ vary from 10 Ω for a 20 MHz crystal to 200 kΩ for 1 kHz crystals.

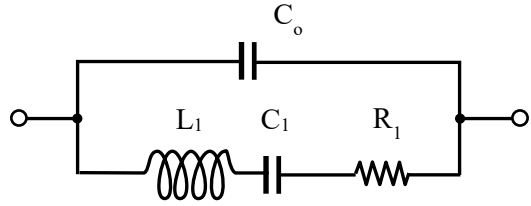

Fig. 32: Equivalent circuit for a quartz crystal

The Q factor of a series LCR resonator is given as $\frac{1}{R}\sqrt{\frac{L}{C}}$ and the frequency as $\frac{1}{2\pi\sqrt{LC}}$. As an example for $C_1$= 24.6 fF, $L_1$ =10.298 mH, R = 16 Ω then the frequency is 9.999450 MHz, and the Q factor is 40438.

The shunt capacitance $C_o$ is the sum of capacitance due to electrodes on the crystal (independent of piezo effects) and also stray capacitance due to its mounting. The value is typically in the range 3 pF to 7 pF. The transfer function of the circuit of Fig. 32 is easily determined from a Spice model like LT Spice. Using the above example and taking $C_o$ as 7 pF Fig. 33 gives the current through the circuit of Fig. 32 for an AC applied peak voltage of 1.5V. The solid line gives the magnitude, and the dotted line gives the phase.

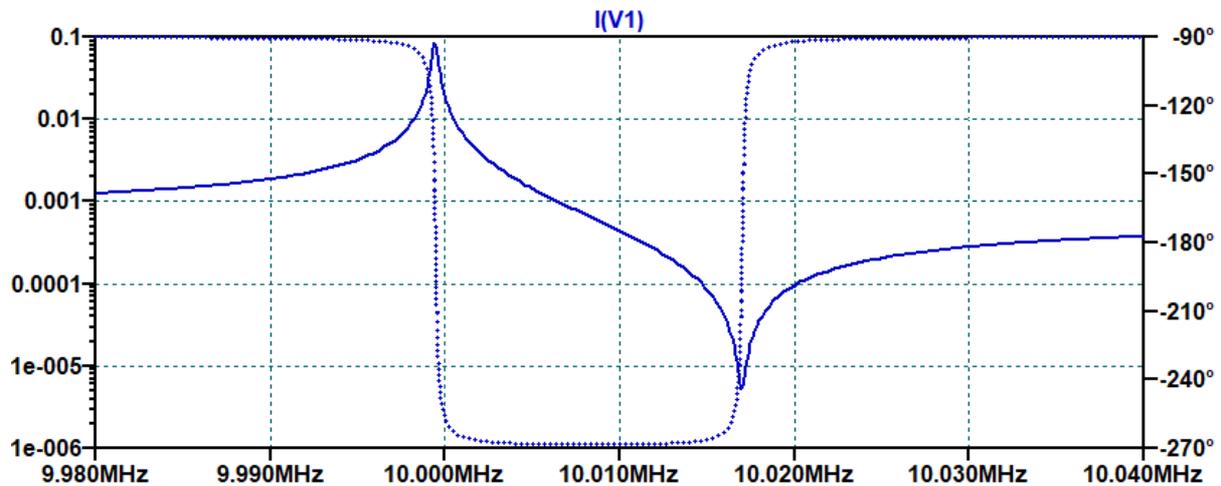

Fig. 33: Frequency response of a 10 MHz quartz crystal

The bandpass peak is at 9.99945 MHz as expected. For frequencies less than this the crystal is capacitive. The band stop dip is at 10.01700 MHz. Between the bandpass peak and the band stop dip the crystal is inductive. The oscillation frequency is set in the inductive region and where the phase changes rapidly which in this case would be at 10.000 MHz.

The oscillator can be implemented by replacing the $L_1$ in Fig. 31 with the crystal. A small tuning range can be achieved by placing a varactor diode in parallel with $C_2$ and $C_3$. The capacitors $C_2$ and $C_3$ load the crystal and make it less sensitive to noise. Their series combination is the load capacitance of the crystal as specified by the manufacturer as is required to obtain the nominal frequency. This would be a Pierce implementation. The other common implementation is using the Colpitts circuit.

### 9.5 Dielectric resonance oscillators

Figure 28 shows a dielectric resonance oscillator (DRO) locked to a 10 MHz master oscillator. It is important to maximise clock stability at a number of important frequencies employed in an accelerator complex. Stability is enhanced by high Q oscillators which invariably have a very small tuneable range. For this reason, one would not choose general wide-band VCOs for critical clocks. Quartz crystal overtones allow high Q oscillators to be made up to about 300 MHz, but above that, other resonators are needed. Surface wave acoustic resonators (SAW) are useful in the range of 200 MHz to 2 GHz. Above this frequency, the high Q resonator most widely employed is the DRO. Their fundamental frequencies are usually in the range 6 GHz to 12 GHz. Using the first sub-harmonic they might provide frequencies down to 3 GHz and higher harmonics are used up to about 50 GHz. Q factors up to $10^5$ are possible. Another technology that can be used in the frequency range 1.5 GHz to 3 GHz is the thin-film bulk acoustic resonator (TFBAR).

The DRO uses a small, carefully shaped, ceramic block, with a very high dielectric constant whose relative permittivity is in the range 30-80. Barium Titanate offers a Q factor of 9000 at 10 GHz and a low-temperature coefficient of +/- 6 ppm/°C [24]. The shape is chosen to ensure that the operating mode is well separated from other modes. The resonance can be modelled as a parallel LCR circuit. It is usually magnetically coupled to a microstrip. Figure 34 illustrates a series scheme where feedback comes from the transistor utilising their nonzero $S_{12}$ at microwave frequencies.

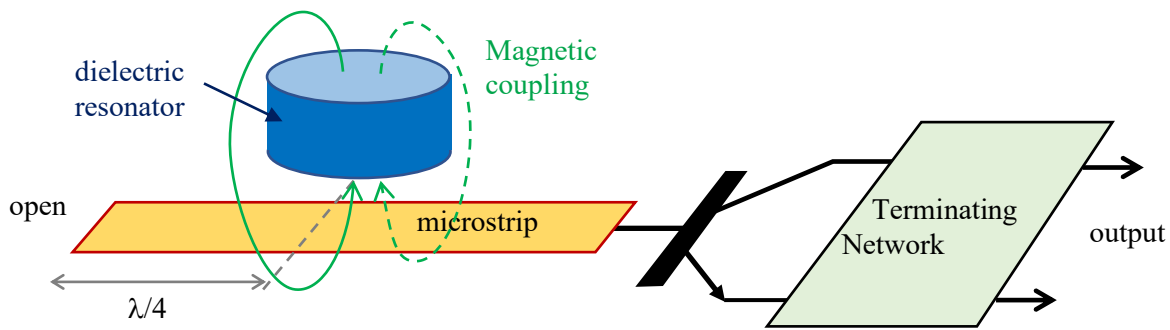

**Fig. 34:** Series dielectric resonator operation assuming a small level of feedback through the transistor.

## 10 Phase noise

Time domain variations in the phase of a signal are known as jitter and are illustrated in Fig. 35 where the red dashed line does not cross zero at the desired times as given by the perfect sine wave in blue. Jitter in the RF signals and power is a problem for accelerators as bunches of charge turn up on time and if phasing is incorrect dispersion is increased.

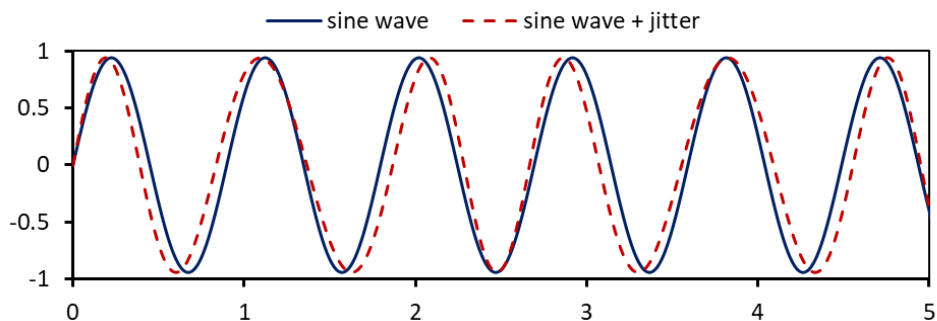

**Fig. 35:** Signal jitter

Amplitude noise is usually less pronounced on RF signals as the oscillator in the source and the power amplifiers run near saturation, also amplitude errors tend not to cause dispersion. In principle, jitter can be measured on an oscilloscope however the speed and accuracy of sampling are invariably insufficient to get useful measurements.

For this reason, jitter is usually examined in the frequency domain where it is called phase noise. Consider a sine wave with a regular oscillation on the phase representing the unwanted phase jitter and a regular oscillation on the amplitude representing unwanted amplitude jitter then

$$V = V_o \{1 + a_n \cos(\omega_n t)\} \cos\{\omega t + \varphi_n \cos(\omega_n t)\}, \qquad (42)$$

where $a_n$ amplitude modulation depth, $\varphi_n$ is the phase modulation depth, $\omega$ is the angular frequency of the signal and for the example to be given, $\omega_n$ is taken as the angular frequency of both the amplitude and the phase noise. Fig. 36 gives the spectral density derived from the Fast Fourier Transforms of Eq. (42) for the two cases of $a_n$ being zero on the left (phase modulation) and $\varphi_n$ being zero on the right (amplitude modulation). The noise frequency was taken as 1 MHz in both cases, the modulation depth $\phi_n$ was taken as $\pi/20$ radians, and the amplitude modulation was taken as 5%. As the maximum modulation for phase is to 180° out of phase and the maximum useful modulation of amplitude is to one, then there is equivalence in this choice of modulation.

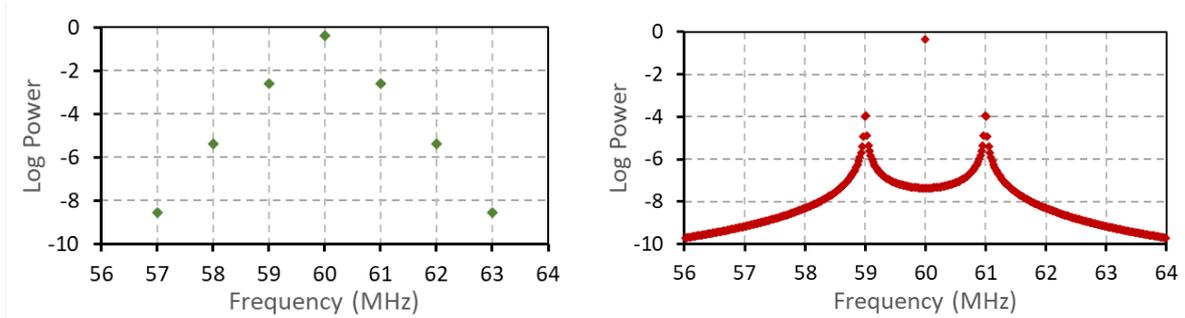

**Fig. 36:** Spectral densities for phase-modulated signal (left) and amplitude-modulated signal (right)

Comparing the highest noise peaks at 59 MHz and 61 MHz between the two plots in Fig. 34 it can be seen that the largest noise contribution is for the phase modulation at $10^{(-2.58)} = 26.3\text{e}^{-4}$ whilst for the amplitude modulation it is only $1.0\text{e}^{-4}$. This indicates that the spectral density preferentially picks out the phase noise. An estimate of jitter can be made by expanding Eq. (42) with $a_n = 0$ and to find the first spectral peak,

$$V = V_o \cos\{\omega t + \varphi_n \cos(\omega_n t)\} = \cos(\omega t)\cos\{\varphi_n \cos(\omega_n t)\} - \sin(\omega t)\sin\{\varphi_n \cos(\omega_n t)\},$$

when the noise is small then $\varphi_n$ is small, so this expression simplifies to,

$$V/V_o \approx \cos(\omega t) - \frac{\varphi_n}{2}\sin[(\omega + \omega_n)t] + \frac{\varphi_n}{2}\sin[(\omega - \omega_n)t], \qquad (43)$$

Noise that modulates the phase at frequency $\omega_n$ is mapped to upper and lower sidebands at $\omega+\omega_n$ and $\omega-\omega_n$. A spectrum analyser gives its output as squares of the Fourier coefficients normalised to power per Hz. The sideband power therefore primarily gives a measure of the phase jitter. Phase noise is usually plotted on just one side of the carried-in units of decibels below the carrier. Figure 37 shows the measured single sideband phase noise (SSB) in dBc per Hz for a DRO locked to a 10 MHz source. Spectrum analysers often have the capability to make this measurement and remove the carrier.

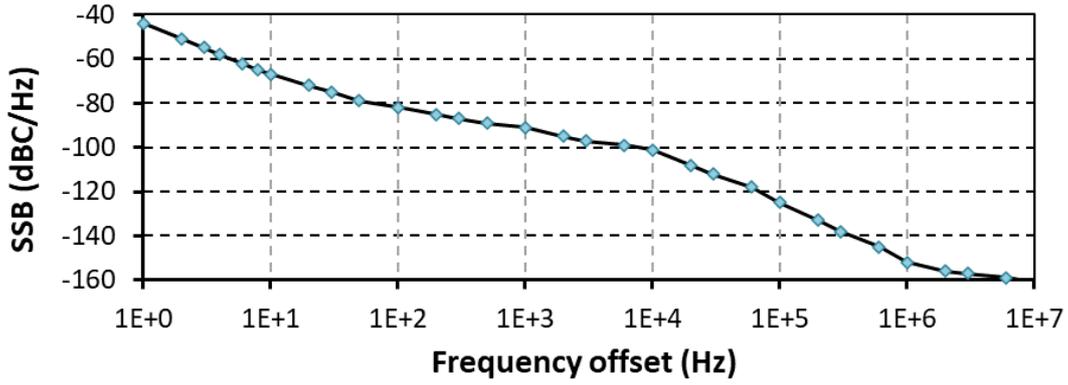

**Fig. 37:** Phase noise plot generated for MITEQ 3.9 GHz DRO locked to a 10 MHz source.

Taking the values given in dBC/Hz as SSB, then the magnitude of the Fourier coefficient is $\sqrt{10^{-SSB/10}}$ so that the peak deviation in Eq. (41) is given as $\varphi_n = 2\sqrt{10^{-SSB/10}}$.
As Eq. (41) just considered a sine wave, then the rms phase deviation $\varphi_{rms} = \varphi_n/\sqrt{2}$ hence, $\varphi_{rms}^2 = 2\times 10^{-SSB/10}$.

If there are a large number of frequencies present in the signal and they are uncorrelated, then they each add a phase error. These phase errors come as components of the Fourier transform so they are sinusoidal. Random phases add as length vectors in a Brownian random motion i.e.

Net distance travelled $= \sum_n \underline{\delta L_i} \approx \sqrt{\sum_n |\delta L_i^2|}$ where $\underline{\delta L_i}$ = vectored step length, hence,

$$\varphi_{rms}^2 = \frac{1}{T}\int_0^T \varphi^2(t)\,dt = 2\int_{1/T}^{\infty} 10^{-SSB/10}\,df, \qquad (42)$$

where $\varphi(t)$ is the time domain jitter and f is the frequency. Note that at low frequencies in Fig. 37 the noise is rising with 3 orders of magnitude per decade hence a $1/f^3$ law applies. This is typical of $1/f$ flicker noise from a transistor being attenuated by a high Q filter. The $1/f^3$ does not continue to zero as this would imply infinite deviations on very long time scales. The curve should become Lorentzian near zero. Applying Eq. (42) to Fig. 37 gives rms jitters of 0.655°, 0.247° and 0.189° when one integrates from 10 MHz to 1 Hz, 10 Hz and 100 Hz respectively.

## 11 Phase-locked loops

A phase-locked loop (PLL) uses a frequency divider in a feedback loop to force a VCO to oscillate at a multiple frequency of an input reference frequency. At low-frequency offsets from the centre frequency, phase noise is constrained to follow the performance of the reference. At high frequencies, the loop filter ceases to be effective, and noise is determined by the performance of the free-running VCO. Figure 38 shows a PLL configuration where the phase comparison is made at the reference frequency. Communication systems typically include further division so that frequencies are compared at a subharmonic of the reference frequency and the VCO can be stepped in smaller intervals than the reference frequency. Non-integer division is also possible by averaging integer divisions, but this increases noise.

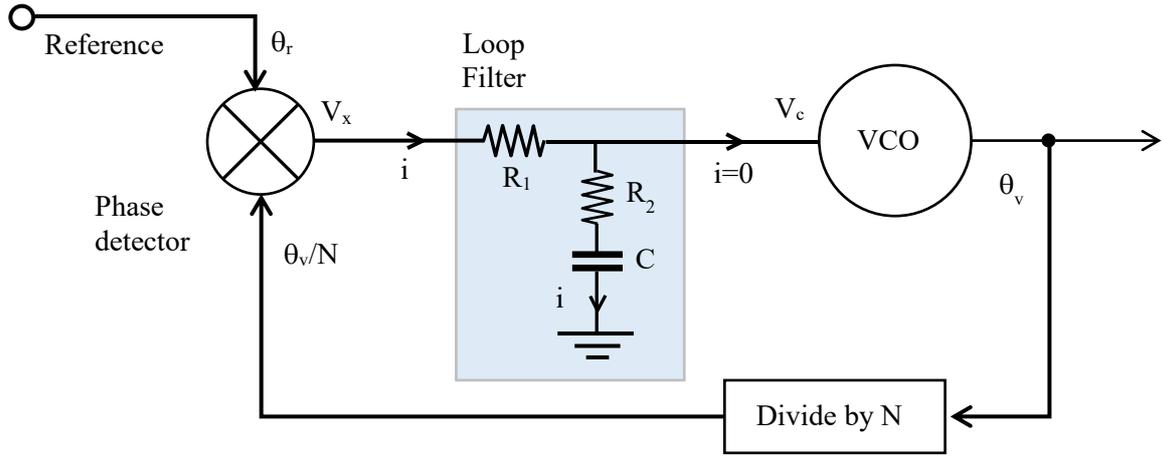

**Fig. 38:** Phase Locked Loop Schematic

The locking performance is determined as follows.

Let the phase of the reference be given as $\theta_r = \omega_r t$ where $\omega_r$ is its frequency.

Let the phase of the VCO be given as $\theta_v = \omega_v t$ where $\omega_v$ is the instantaneous VCO frequency.

The RF output from the VCO gets divided by an integer N and this becomes an input to the mixer. The mixer compares the reference frequency with a signal that is very close to this frequency hence the output from the mixer is a slowly varying voltage $V_x$ given by:

$$V_x = k_x \left( \theta_r - \theta_v / N \right). \tag{43}$$

Assuming the VCO in Fig. 38 has infinite input impedance, then the input and output voltages for the loop filter are determined by the equations:

$$V_x = i(R_1 + R_2) + (1/C) \int i \, dt, \tag{44}$$

$$V_c = iR_2 + (1/C) \int i \, dt, \tag{45}$$

where the current and component values are defined in the figure.

Taking the output frequency of the VCO as a linear function of the input voltage $V_c$ then:

$$\frac{d\theta_v}{dt} = k_v V_c. \tag{46}$$

Take the Laplace transforms of Eqs. (43)-(46) to give:

$$\tilde{V}_x = k_x \left( \tilde{\theta}_r - \tilde{\theta}_v / N \right), \tag{47}$$

$$\tilde{V}_x = \tilde{i} \left( R_1 + R_2 + \frac{1}{sC} \right), \tag{48}$$

$$\tilde{V}_c = \tilde{i} \left( R_2 + \frac{1}{sC} \right), \tag{49}$$

$$s \tilde{\theta}_v = k_v \tilde{V}_c. \tag{50}$$

Eliminating $\tilde{v}_x$, $\tilde{v}_c$, and $\tilde{i}$ gives

$$\frac{s}{k_v}\tilde{\theta}_v = \frac{1+sCR_2}{1+sC(R_1+R_2)}k_x\left(\tilde{\theta}_r - \frac{\tilde{\theta}_v}{N}\right). \tag{51}$$

Hence

$$\frac{s}{k_v}\tilde{\theta}_v = \frac{1+sCR_2}{1+sC(R_1+R_2)}k_x\left(\tilde{\theta}_r - \frac{\tilde{\theta}_v}{N}\right), \tag{52}$$

gives

$$\tilde{\theta}_v = N\frac{1+sCR_2}{1+s\left(\frac{N}{k_x k_r}+CR_2\right)+s^2\left\{C(R_1+R_2)\frac{N}{k_x k_r}\right\}}\tilde{\theta}_r. \tag{53}$$

This is a second-order system as the denominator has two poles in the negative half-plane as all coefficients in the quadratic equation are positive. Steady-state response is determined by taking the limit of s going to zero so that:

$$\tilde{\theta}_v = N\tilde{\theta}_r \quad \text{for} \quad s=0.$$

Hence after a long time, the VCO has exactly N times the reference frequency.

The period of oscillation associated with the VCO chasing N times the reference frequency is given as

$$T = 2\pi\sqrt{C(R_1+R_2)\frac{N}{k_x k_v}}.$$

The damping factor for the quadratic equation in the denominator is

$$\gamma = 0.5\left(R_2 C + \frac{N}{k_x k_v}\right)\sqrt{\frac{k_x k_v}{NC(R_1+R_2)}}.$$

Not included in this model is the time delay for the loop. This limits the gain and hence the minimum period T. Once T has been made as small as practical, the damping factor can be set. Typically, γ is taken as about 0.7 to get an optimally flat frequency transfer characteristic.

The above analysis for PLL operation would not normally be used as suppliers like Analog Devices provide software for design using their components. Their specific software is called ADI SimPLL and an example of its output is given in Fig. 39. In this example the period T has been set to about 20 μs which is 50 kHz. At higher frequencies, the PLL makes no reduction of the free-running VCO jitter towards the phase noise level of the reference.

## 12 Final remarks

If this is your first course on RF electronics then hopefully it is the beginning, not the end. As an accelerator engineer, it is unlikely you will ever need to design your own chips or microwave components. You may work with PCBs or perhaps just with connectorized components. Your task is likely to be that of a systems engineer where subsystems must be integrated to give the best possible overall performance. This means you need an excellent understanding of what components are available, what they do, how to integrate them and how to test overall performance. Your main learning resource is likely to be the manufacturer's datasheets and application notes. This course has barely scratched the surface of the range of components available.

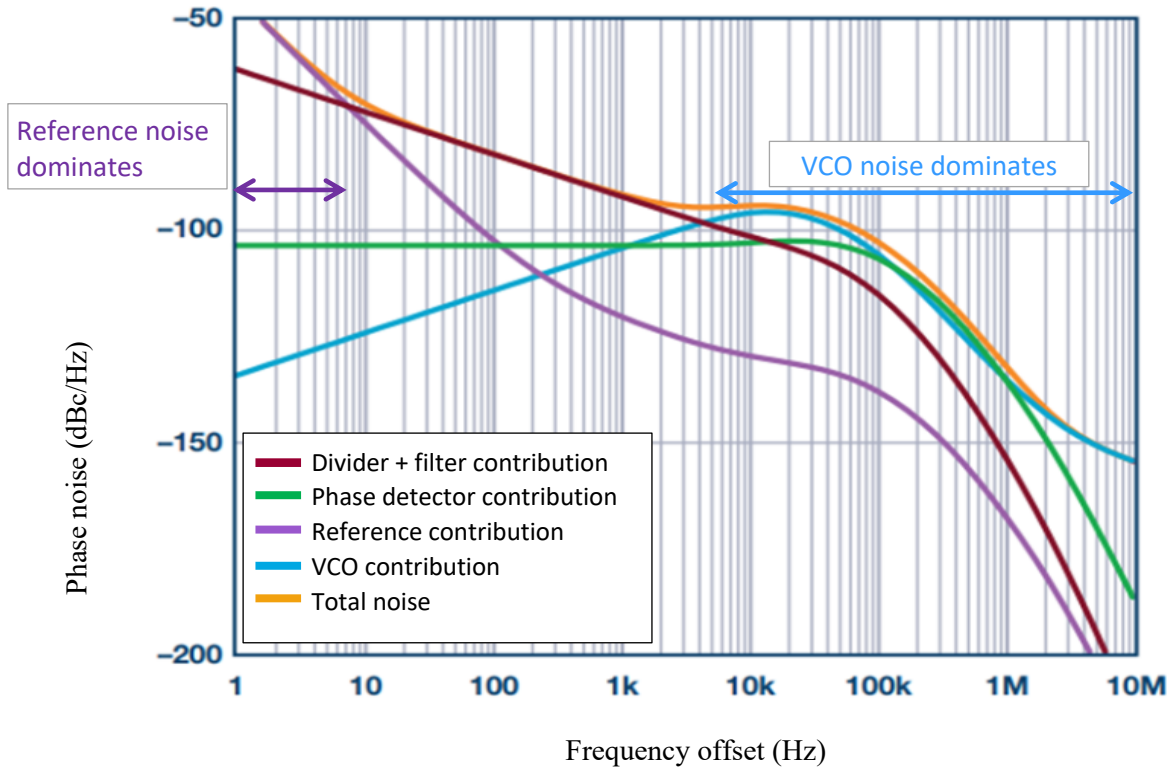

**Fig. 39:** Example output from Analog Devices software ADI SimPLL.